\documentclass[reprint,pre]{revtex4-2}
\usepackage{dcolumn}
\usepackage{bm}

\usepackage[utf8]{inputenc}
\usepackage[T1]{fontenc}
\usepackage{mathptmx}
\usepackage{etoolbox}
\usepackage{amsmath}
\DeclareMathOperator\erf{erf}

\usepackage{graphicx}    

\usepackage{siunitx}
\usepackage{xcolor}

\usepackage{latexsym}
\usepackage{wasysym}

\usepackage{float} 

\usepackage{tabularx}

\usepackage{amssymb}

\usepackage{ulem}

\makeatletter
\def\@email#1#2{%
 \endgroup
 \patchcmd{\titleblock@produce}
  {\frontmatter@RRAPformat}
  {\frontmatter@RRAPformat{\produce@RRAP{*#1\href{mailto:#2}{#2}}}\frontmatter@RRAPformat}
  {}{}
}%
\makeatother

\begin{document}

\title{Molecular rotors for \textit{in situ} local viscosity mapping in microfluidic chips}

\author{Dharshana Nalatamby}
\author{Florence Gibouin}

\affiliation{Univ. Bordeaux, CNRS, Solvay, LOF, Pessac, F-33600, France}

\author{Javier Ord{\'o}{\~n}ez-Hern{\'a}ndez}

\affiliation{Facultad de Qu{\'i}mica, Departamento de Qu{\'i}mica Org{\'a}nica, Universidad Nacional Aut{\'o}noma de M{\'e}xico, 04510, Ciudad de M{\'e}xico, M{\'e}xico}

\author{Julien Renaudeau}
\author{G{\'e}rald Clisson}

\affiliation{Univ. Bordeaux, CNRS, Solvay, LOF, Pessac, F-33600, France}
\author{Norberto Farf{\'a}n}

\affiliation{Facultad de Qu{\'i}mica, Departamento de Qu{\'i}mica Org{\'a}nica, Universidad Nacional Aut{\'o}noma de M{\'e}xico, 04510, Ciudad de M{\'e}xico, M{\'e}xico}

\author{Pierre Lidon*}
\email[]{pierre.lidon@u-bordeaux.fr}

\author{Yaocihuatl Medina-Gonz{\'a}lez*}
\email[]{yaocihuatl.medina-gonzalez@u-bordeaux.fr}

\affiliation{Univ. Bordeaux, CNRS, Solvay, LOF, Pessac, F-33600, France}

\date{\today}


\begin{abstract}
In numerous industrial processes involving fluids, viscosity is a determinant factor for reaction rates, flows, drying, mixing, etc. Its importance is even more determinant for phenomena observed are at the micro- and nano- scales as in nanopores or in micro and nanochannels for instance. However, despite notable progresses of the techniques used in microrheology in recent years, the quantification, mapping and study of viscosity at small scales remains challenging. Fluorescent molecular rotors are molecules whose fluorescence properties are sensitive to local viscosity: they thus allow to obtain viscosity maps by using fluorescence microscopes. While they are well-known as contrast agents in bioimaging, their use for quantitative measurements remains scarce. This paper is devoted to the use of such molecules to perform quantitative, \textit{in situ} and local measurements of viscosity in heterogeneous microfluidic flows. The technique is first validated in the well-controlled situation of a microfluidic co-flow, where two streams mix through transverse diffusion. Then, a more complex situation of mixing in passive micromixers is considered and mixing efficiency is characterized and quantified. The methodology developed in this study thus opens a new path for viscosity characterization in confined, heterogeneous  and complex systems.
\end{abstract}

\maketitle

\section{Introduction}
\label{sec:1_intro}

In any process involving fluid flows, viscosity is a key control parameter. Viscosimeters are the most common tool to measure this quantity, but their use requires great care to avoid artifacts \cite{Ewoldt2015}. More importantly, they only measure a viscosity averaged over macroscopic volumes and are difficult to implement inline, which strongly limits their application in industrial processes or in confined systems, or available in very limited volume. The impossibility of obtaining local data also impedes their use with complex flows involving spatial heterogeneity (e.g., inpaint manufacturing, food processing or biomedical applications) and at small scales (e.g., in nano-, micro- or millifluidics multi-phase flows or in porous media, with applications in enhanced oil recovery, and catalysis among others). Designing tools for small-scale viscosity measurements is thus an important stake for fundamental studies as well as industrial and medical applications \cite{Arnes1999}.  

Microfluidics has been widely used as an essential tool in numerous applications such as high-throughput screening in the research and development domain~\cite{Dunn2000} and chemical reactions analysis~\cite{Losey2001} and give interesting opportunities for viscosity measurements. For instance, measuring pressure drop along a microchannel at an applied flow rate allows the determination of viscosity averaged on tiny volumes, below $\SI{1}{\micro\liter}$~\cite{Kang2005,Srivastava2005}. More local approaches require the characterization of the flow profile by introducing fluorescent tracers in the fluid~\cite{Galambos1998,Guillot2006}, but are limited to simple configurations and remain averaged over mesoscopic scales.

Fluorescent molecular rotors (FMR) offer a direct path for local viscosity measurements. These are fluorescent molecules whose conventional fluorescent relaxation after photoexcitation is in competition with a non-radiative mechanism involving the rotation of a molecular bond~\cite{Haidekker2016}. This motion is hindered by the local micro-viscosity of the environment~\cite{Strickler1962, Haidekker2010}, with higher micro-viscosity leading to increased quantum yield, thus more intense fluorescence with longer lifetime~\cite{Kung1989}. While the precise relationship between micro-viscosity and the usual viscosity remains unclear, they are directly related in molecular fluids and after preliminary calibration, fluorescent measurements can be used to retrieve local viscosity.

FMR have been acknowledged as excellent local viscosity probes with real-time response and high spatial resolution~\cite{Benninger2005,Haidekker2007, Ponjavic2015} and are, for instance, used as contrast agents in cells bioimaging~\cite{Kuimova2008,Wu-Kuimova2013}. However, their use for quantitative characterization in other contexts remains scarce~\cite{Bunton_HeleShaw,Garg2020}. More particularly in microfluidic context, regular and confocal fluorescence lifetime imaging (of FMR and other fluorophores) have been proved to be powerful tools for mapping viscosity in complex, three-dimensional flows~\cite{redford_2005,benninger_2006,srisaart_2008,casadevallisolvas_2010} but they were performed with a sophisticated lifetime measurement setup operating directly in time domain, which requires a pulsed laser source, and are thus difficult to be applied by non-specialists.

In this work, a FMR was synthesized and used for fluorescence lifetime imaging in the frequency domain, using a commercial apparatus. The measurements were performed in well controlled experiments of purely diffusive mixing of two miscible streams co-flowing in a simple Y-mixer microfluidic channel; the obtained results were satisfactorily compared with existing models and results from the literature~\cite{Salmon2005,Dambrine2009}. After the validation of the method in this simple case, it was finally used to assess the mixing efficiency of a passive micromixer. These results validate the potential of the use of FMR for quantitative and local measurements of viscosity in microfluidic flows and open perspectives for fluid characterization.

\section{Material and Methods}

\subsection{BODIPY-2-OH synthesis and characterization}

\begin{figure}[!htb]
    \centering
    \includegraphics[scale=0.5]{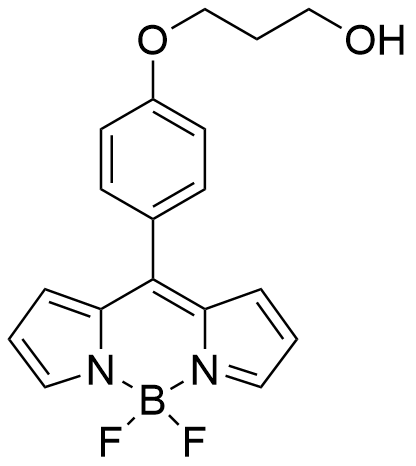}
    \caption{Chemical structure of BODIPY-2-OH synthesized in this work.}
    \label{bodipy}
\end{figure}

The viscosity-sensitive fluorescent boron-dipyrromethene (BODIPY)-based probe,  BODIPY-2-OH (see Figure~\ref{bodipy}), was developed specifically  for this study. Based on previously reported rotors, \cite{Kuimova2013} BODIPY-2-OH was designed to ensure its viscosity-sensitive characteristics. Adding an OH group and a short chain to the BODIPY core allowed to improve solubility in polar solvents.~\cite{prlj_2017,kee_2005,refMexico} Details on its synthesis are provided in the Supporting Information. In this structure, the alkoxyphenyl unit is an electron donor group in conjugation with the BODIPY core, which is an electron acceptor group. After photoexcitation, relaxation of the molecule to its ground state occurs through conventional fluorescent photoemission, accompanied by a rotation around the single bond linking alkoxyphenyl and BODIPY groups~\cite{alamiry_2008}. The radiative relaxation rate is affected by the refractive index of the surrounding medium, while the non-radiative relaxation rate depends on the free volume of the micro-environment, related to the viscosity~\cite{Strickler1962,Haidekker2010}.

Mixtures of DMSO and glycerol were chosen as working fluids in this article, as their viscosity significantly vary with their composition~\cite{Angulo2016} and because BODIPY-2-OH is very soluble in these solvents. The volume fraction of glycerol in analyzed solutions was controlled, and the concentration of BODIPY-2-OH was kept constant at $\SI{e-5}{\mol\per\litre}$ in all the experiments.

\subsubsection{Synthesis of BODIPY-2-OH}

All starting materials were purchased from Sigma-Aldrich and used without further purification. Solvents were dried by standard methods or distilled prior to use. Reactions were monitored by thin-layer chromatography on pre-coated silica gel plates (ALUGRAM SIL G/UV254) and revealed by exposure to a UV lamp (254 nm). Infrared spectra were obtained using a Perkin-Elmer Spectrum 400 FT-IR/FT-FIR spectrophotometer, wavelength is reported in $\SI{}{\per\centi\meter}$. $^1$H, $^{13}$C, $^{11}$B and $^{19}$F NMR spectra were recorded using Varian Unity Inova 300, JEOL ECA 400 spectrometers, chemical shifts ($\delta$/ppm) are reported relative to $\mathrm{Si(CH_3)_4}$, $\mathrm{CDCl_3}$, $\mathrm{BF_3OEt_2}$, and $\mathrm{CDCl_3}$. High-resolution mass spectrometry (HR-MS) spectra were acquired with an Agilent Technologies ESI TOF spectrometer.

\begin{figure*}[!htb]
    \centering
    \includegraphics[scale=0.3]{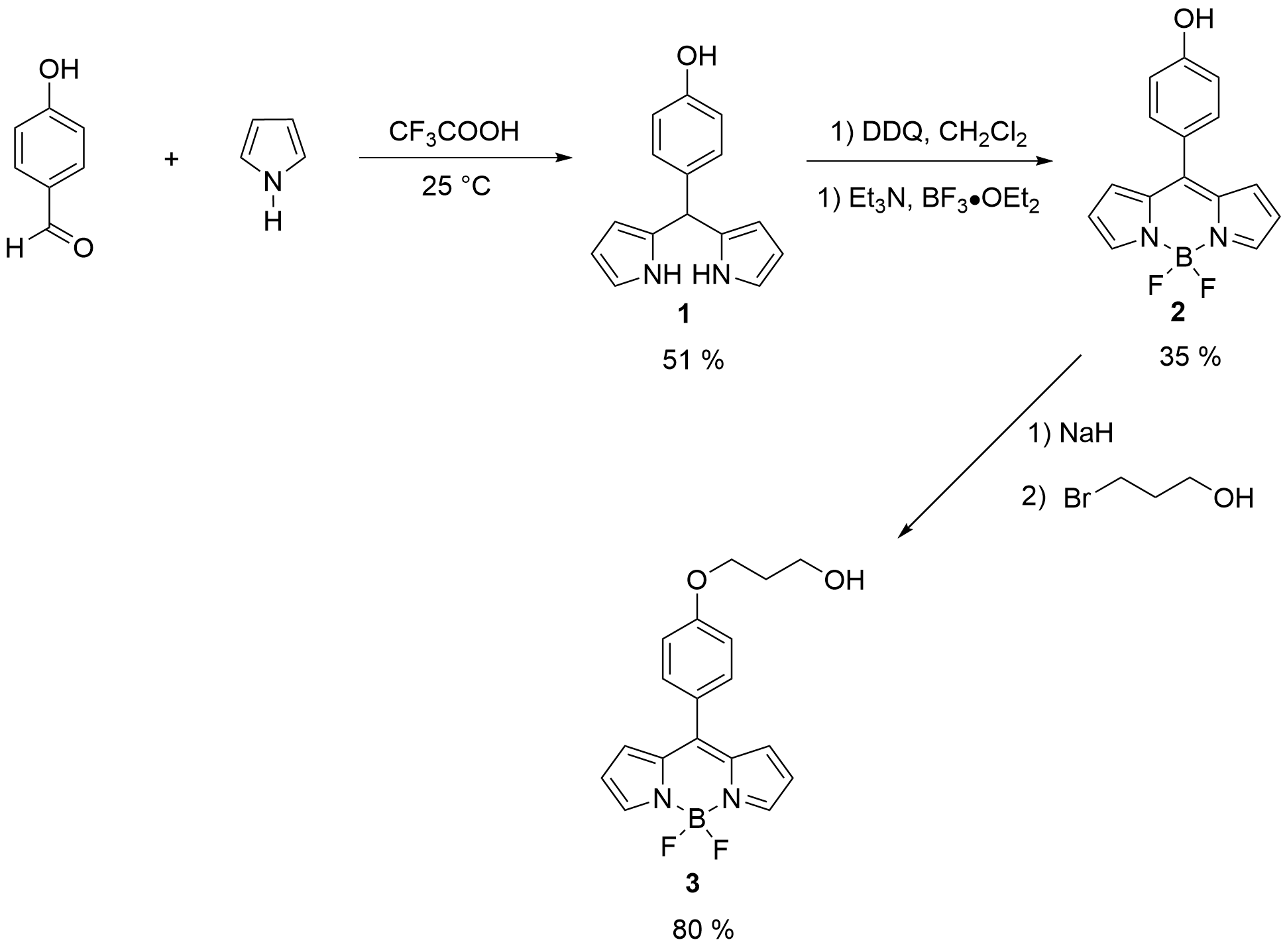}
    \caption{Synthesis of BODIPY-2-OH.}
    \label{scheme1}
\end{figure*}

The \textit{meso-}substituted BODIPY-2-OH was prepared by the synthetic route shown in Figure~\ref{scheme1}. Compound (1), a dipyrromethane ($\SI{51}{\percent}$ yield) derivative was synthesized from the condensation reaction between the corresponding \textit{p}-hydroxibenzaldehyde and ten equivalents of pyrrole in the presence of a catalytic amount of $\mathrm{CF_3COOH}$. \cite{refMexico} The oxidation of compound (1) with 2,3-dichloro-5,6-dicyano-1,4- benzoquinone (DDQ)  followed by a complexation reaction $\mathrm{BF_3.OEt_2}$ to give Compound (2) ($\SI{35}{\percent}$ yield).  Compound (2) was dissolved in tetrahydrofuran (THF) and sodium hydride was added. After 30 minutes, 3-bromo-1-propanol was added to the reaction mixture to obtain BODIPY-2-OH ($\SI{80}{\percent}$ yield). The final product was characterized by spectroscopic techniques such as $^1$H, $^{13}$C, $^{11}$B, $^{19}$F NMR, IR and HR-MS. Detailed information on the synthesis and NMR spectra of BODIPY-2-OH can be found in the Supporting Information.\\ 

\subsubsection{Photophysical characterization of BODIPY-2-OH}

The absorption and emission spectra of BODIPY-2-OH were  measured respectively using an Agilent Technologies Cary UV-Visible Spectrophotometer and an Agilent Technologies Cary Eclipse Fluorescence Spectrophotometer. BODIPY molecules are susceptible to photodegradation after constant irradiation~\cite{Rybczynski2021}: this process is slow enough to be negligible over the duration of our experiments, but samples were stored in amber vials to limit bleaching by ambient light. Single-use cuvettes (BRAND GMBH $\SI{70}{\micro\liter}$ UV-Cuvette micro) were used for excitation and emission spectra measurements.

\begin{figure}[!htb]
    \centering
    \includegraphics[height=5.5cm]
    {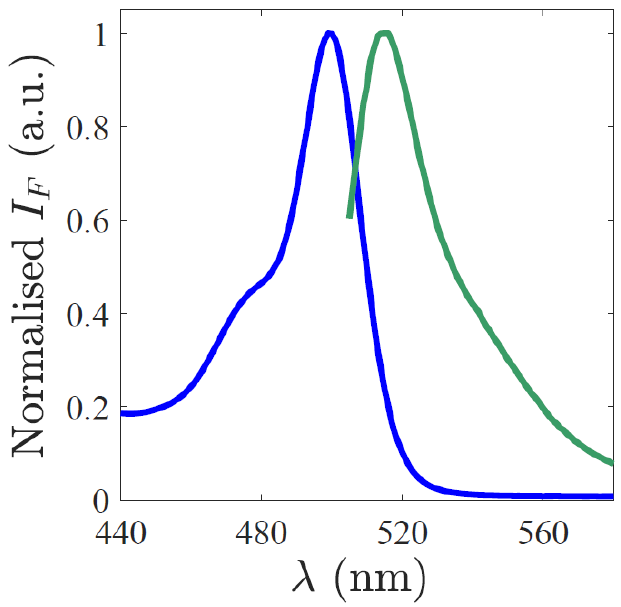}
    \caption{Absorption (blue line) and emission spectra (green line) of BODIPY-2-OH in glycerol at a concentration $c=\SI{e-5}{\mol\per\litre}$. Maximum excitation and emission wavelengths of BODIPY-2-OH in glycerol are respectively $\SI{500}{\nano\meter}$ and $\SI{515}{\nano\meter}$.}
    \label{BODIPY_peak}
\end{figure}

The obtained spectra for solutions of BODIPY-2-OH in glycerol at a concentration $c=\SI{e-5}{\mol\per\litre}$ are depicted in Figure~\ref{BODIPY_peak}. The maximum excitation and emission wavelengths were respectively found to be $\SI{500}{\nano\meter}$ and $\SI{515}{\nano\meter}$. These spectra are similar to those previously reported for BODIPY-based molecular rotors, which proves that the fluorescence characteristics are not significantly affected by the different groups attached to the BODIPY base \cite{Wu-Kuimova2013,Kuimova2013}. We verified that for the range of concentration used in this study, there is no sign of aggregation of the rotor, that could be observed by qualitative changes in absorption and emission spectra.

\subsection{Rotor response to viscosity}

The fluorescence response of a given FMR to viscosity depends on the nature of the solvent environment and has to be carefully calibrated in order to map viscosity quantitatively~\cite{Akers2005, JournalReviewHaidekker2010}. Hence, calibration curves were established by measuring BODIPY-2-OH response in mixtures of known viscosity prior to the use in microdevices. More precisely, two parameters were characterized: emission intensity under steady illumination and fluorescence lifetime, which quantifies the average lifetime of the excited state~\cite{Kafle2020}. Intensity is a convenient parameter to acquire with spectrophotometers or fluorescence microscopes. However, fluorescence intensity does not depend only on the solvent viscosity but also on local dye concentration and excitation intensity, which is affected by the whole optical path before reaching the sample. Careful and tedious calibrations would thus be required to go beyond simple qualitative observations by using fluorescence intensity. On the contrary, while more subtle to measure, fluorescence lifetime is only determined by the microviscosity of the dye and allows for a more straightforward interpretation: this parameter was thus preferred in this study~\cite{Ponjavic2015, Kuimova2008, Kafle2020}.

In order to calibrate the response of BODIPY-2-OH to viscosity, solutions of the rotor in DMSO/glycerol mixtures were prepared from an initial dye stock solution in DMSO. Viscosity of the mixtures were tuned by changing the ratio of DMSO and glycerol and the concentration of BODIPY-2-OH was kept constant at $c=\SI{e-5}{\mol\per\litre}$ to avoid any change in fluorescence intensity due to dye concentration. 

\subsubsection{Bulk viscosity measurements}

The viscosity $\eta$ of the different DMSO/glycerol mixtures were first determined by rheometry. Tests were performed using a Kinexus Ultra+ rheometer (Netzsch) with a Double-Gap geometry (DG24/27 SS CUP) for samples of low viscosity (below $\SI{15}{\milli\pascal\second}$) and $\SI{1}{\degree}$, $\SI{60}{\milli\meter}$-diameter, cone-plate geometry for viscous samples (above $\SI{15}{\milli\pascal\second}$). The experimental temperature was controlled by a Peltier module and set to $T=\SI{25}{\celsius}$. The shear viscosity of every sample was measured by successively applying shear rates of $1$, $10$, $100$, and $\SI{1000}{\per\second}$. Measurements were taken for $\SI{60}{\second}$ for each shear rate with an acquisition rate of 1 point/s and were repeated thrice to ensure good repeatability. All samples displayed Newtonian rheology, and the final viscosity value was taken as an average over all the applied shear rates. Calibration curves relating glycerol volume fraction, glycerol molar concentration, and viscosity of the mixtures were established and fitted by exponential evolutions. Results are provided in Supporting Information.

\subsubsection{Fluorescence intensity measurements}

Absorption and emission spectra of BODIPY-2-OH in the different DMSO/glycerol mixtures were recorded using the previously described protocol. No changes of maximum of absorption ($\lambda_\text{abs} = \SI{500}{\nano\meter}$) and of emission ($\lambda_\text{abs} = \SI{515}{\nano\meter}$) were observed. Maximum of fluorescence emission spectra were recorded in order to quantify the fluorescence intensity.

\subsubsection{Fluorescence lifetime measurements}

Fluorescence lifetimes of BODIPY-2-OH in the different DMSO/glycerol mixtures were measured using a Fluorescence Lifetime Imaging Microscope (FLIM). FLIM technique is a specific case of fluorescence microscopy, enabling the measurement of spatially resolved fluorescence lifetime in heterogeneous samples. It has been, in particular, used with FMR to map qualitative changes of viscosity in bioimaging~\cite{Kuimova2008,Wu-Kuimova2013,Levitt2009}. It is a wide-field method operating in the frequency domain based on a regular setup of fluorescence microscopy with a continuously modulated excitation source. By modifying the phase of the exciting light, the fluorescence lifetime is then calculated for every pixel from the local phase shift of the fluorescence emission~\cite{Suhling2005, Munster2005}. Provided an initial reference has been acquired to account for instrumental behavior, FLIM allows to map fluorescence lifetime over the field of view of the microscope within a few seconds~\cite{Suhling2007}. 

The FLIM experiments presented in this paper were carried out with a LIFA (Lambert Instruments FLIM Attachment) device mounted on an Olympus IX71 inverted microscope. The image acquisitions were carried out at LED modulation frequency of $\SI{40}{\mega\hertz}$ with $12$ acquisition phases and $1\times$ CCD gain. A $\SI{10}{\micro\mole\per\liter}$ fluorescein solution at buffered $\mathrm{p}H=10$ with a tabulated lifetime of $\SI{4.02}{\nano\second}$ was taken as reference. For lifetime calibration, DMSO-glycerol solutions containing FMR at a fixed concentration of $\SI{e-5}{\mole\per\liter}$ were held in cavity slides with glass coverslips used to obtain a flat layer of liquid with even thickness. The samples were illuminated with a LED beam at $\SI{451}{\nano\meter}$ through the dry microscope objectives lenses (10X or 20X Olympus), using a dichroic mirror. The fluorescence emission of the molecular rotor was collected by the same objective and transmitted to the cooled detector after passing through an Olympus 440x-$\SI{490}{\nano\meter}$ long pass filter. The retrieved data were then analyzed using a Matlab application developed by our group. 

Spatial resolution of the measurement is limited by the imaging setup: for instance, in our setup, pixel size for a 20X objective lens corresponds to a distance of $\SI{1.14}{\micro\meter}$. Time resolution was not a question in this study as we only considered steady flows. Yet, it can be specified that acquisition of the full field of the microscope takes typically a few seconds for maximum resolution on lifetime, and can be reduced by acquiring smaller portion of the field of view or decreasing the accuracy. The employed FLIM is adapted for measurement of lifetimes above $\SI{1}{\nano\second}$ with maximum accuracy of about $\SI{0.1}{\nano\second}$.

Signal analysis to retrieve lifetime assumes that fluorescence decay of the MR is monoexponential, with a single characteristic time. It is possible to test this hypothesis by looking at so-called polar plots~\cite{leray_2012}. In the different calibrations and experiments, such verification was performed: while multiexponential decays (e.g. due to autofluorescence of the chip material) cannot be fully ruled out, no clear evidence of such effect was observed.

\subsection{Microfluidics}

\begin{figure*}[!htb]
\centering
\parbox{0.4\textwidth}{\includegraphics[height=2.5cm]{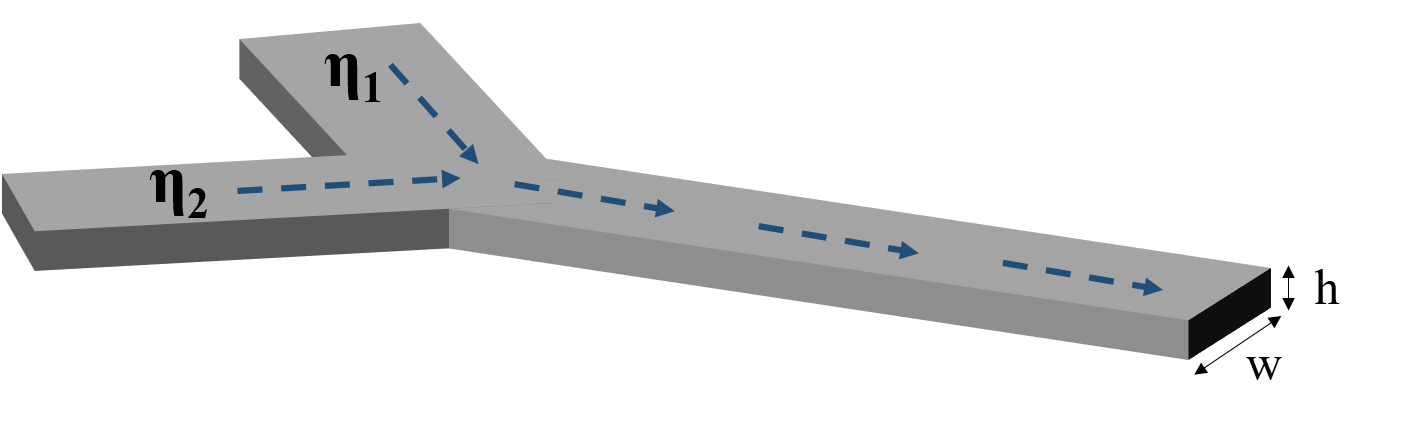}}%
\qquad
\begin{minipage}{0.4\textwidth}%
\includegraphics[height=4.5cm]{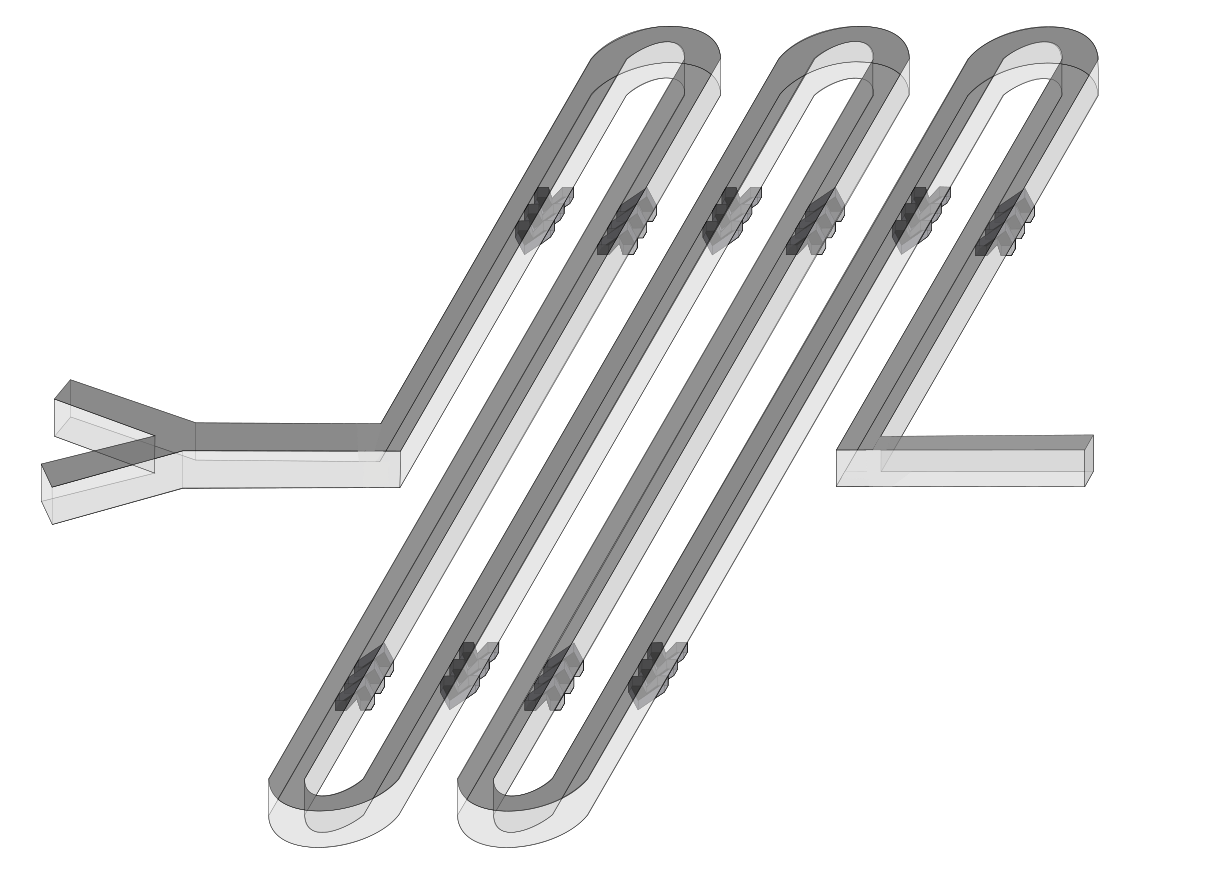}
\end{minipage}%
\caption{(a) Schematics of a  microfluidic Y-junction of height $h=\SI{40}{\micro\meter}$ and width $w=\SI{500}{\micro\meter}$. The mixing of two fluids flowing side by side in the microchannel occurs through transverse molecular diffusion. (b) Schematics of the microfluidic Y-junction with a staggered herringbone passive mixer (SHM) used in this work.} 
\label{fig:chip_schematics}
\end{figure*}

As a proof of principle of quantitative measurements of viscosity using FMR with the proposed setup, two microfluidic configurations involving heterogeneous flows were studied. First, the situation of diffusive mixing of two streams in a simple microfluidic Y-junction, schematized in Figure~\ref{fig:chip_schematics}(a), was considered as it is well controlled and characterized in the literature. Then, the more complex situation of the mixing of two incoming streams in a Y-junction with staggered-herringbone passive micromixers (SHM) was considered, as being qualitatively understood and of practical interest in microfluidic applications~\cite{Stroock2002}. The corresponding chip designs are depicted in Figure~\ref{fig:chip_schematics}(b).

\subsubsection{Microfabrication}

For fabrication of the simple Y-mixer chip, a mixture of $\SI{95}{\percent}$ Polyethylene Glycol Diacrylate (PEGDA)-250 and $\SI{5}{\percent}$ of photoinitiator, 2-hydroxy-2-methylpropiophenone was prepared in advance before being injected by capillarity into the interstitial space between a glass slide and a 2-level negative photoresist mold (SU-8 3050, MicroChem). This configuration was later exposed under the UV lamp of an aligner for $\SI{1.2}{\second}$ (power of UV mercury vapor lamp is $\SI{35}{\milli\watt\per\centi\meter\squared}$ at $\SI{365}{\nano\meter}$). The polymerized PEGDA film was then attached to a silanized glass slide to seal the microchannel. \cite{theseCamille,Decock2018} PEGDA chip was chosen for this experiment because of its fast microfabrication method, which only takes around 5 minutes to make a functional chip\cite{Decock2018}. The dimensions of the main microchannel in Figure~\ref{fig:chip_schematics}(a) were measured after photolithography of the SU-8 mold using a Sensofar Non-contact 3D Optical Profiler. Height and width of the channel were respectively $h=\SI{40}{\micro\meter}$, and the width of the channel, $w=\SI{500}{\micro\meter}$. 

For fabrication of the chip including SHM, A 2-level PDMS microfluidic chip was made using a negative photoresist mold (SU-8 3050, MicroChem) with classic soft lithography techniques (Figure~\ref{fig:chip_schematics}(b)). A glass slide was sealed to the PDMS microfluidic chip after undergoing plasma treatment for 2 minutes. This step was necessary to ensure covalent anchoring of the PDMS block to the glass slide \cite{Whitesides2002}. The chip was then placed in an oven at $\SI{65}{\celsius}$ for 10 minutes to strengthen the seal. The height of the channel without the micromixers is $\SI{31}{\micro\meter}$ and the part with micromixers is $\SI{37}{\micro\meter}$ whereas the width of the channel is $\SI{650}{\micro\meter}$.  These dimensions were measured with the Sensofar Non-contact 3D Optical Profiler with an interferometry acquisition setting (objective lens: 10X Nikon DI). Groups of three SHM grooves occupying a length of $\SI{1.95}{\milli\meter}$ were separated by free intervals of length $\SI{9.40}{\milli\meter}$. The distance from the beginning of the Y-junction to the first group of SHM was about $\SI{7620}{\micro\meter}$. 

\subsubsection{Flow control}

In all experiments (Y-mixer and SHM), mixtures of glycerol and DMSO of different viscosities were injected at flow rates imposed with a neMESYS syringe pump into the two entrance sleeves of the chip. The concentration of BODIPY-2-OH in all solutions was kept constant at $\SI{e-5}{\mol\per\litre}$. Before starting microfluidic experiments, an aqueous solution of BODIPY-2-OH was injected into chips to verify the absence of adsorption or permeation of the molecule into the PDMS matrix. Calibration of the dye response to viscosity was also performed \textit{in situ} by measuring lifetimes in the entrance sleeves, where mixtures have a known composition prior to any mixing. In order to image the chip, a $x-y$  microactuator was used to move the chip above the microscope objective of the microscope, and images were taken along the main microchannel. The mapping of fluorescence lifetime was subsequently carried out using the LIFA-FLIM, and fluorescence data post-processing led to the mapping of viscosity in the microchannel.

\section{Results} 

\subsection{Response of BODIPY-2-OH to viscosity}

\begin{figure*}[!htb]
\centering
\parbox{0.4\textwidth}{\includegraphics[width=0.4\textwidth]{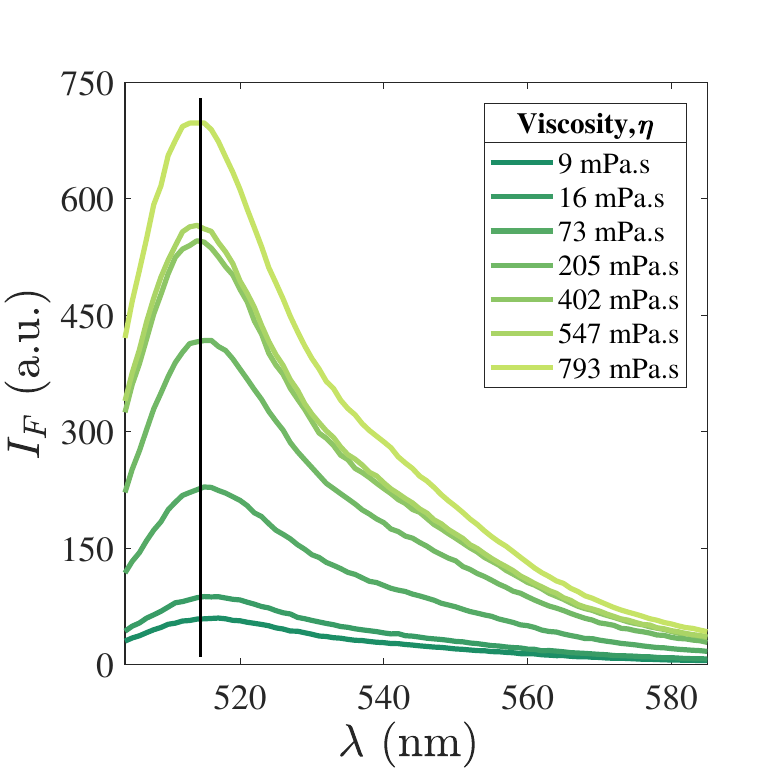}}%
\qquad
\begin{minipage}{0.4\textwidth}%
\includegraphics[width=\textwidth]{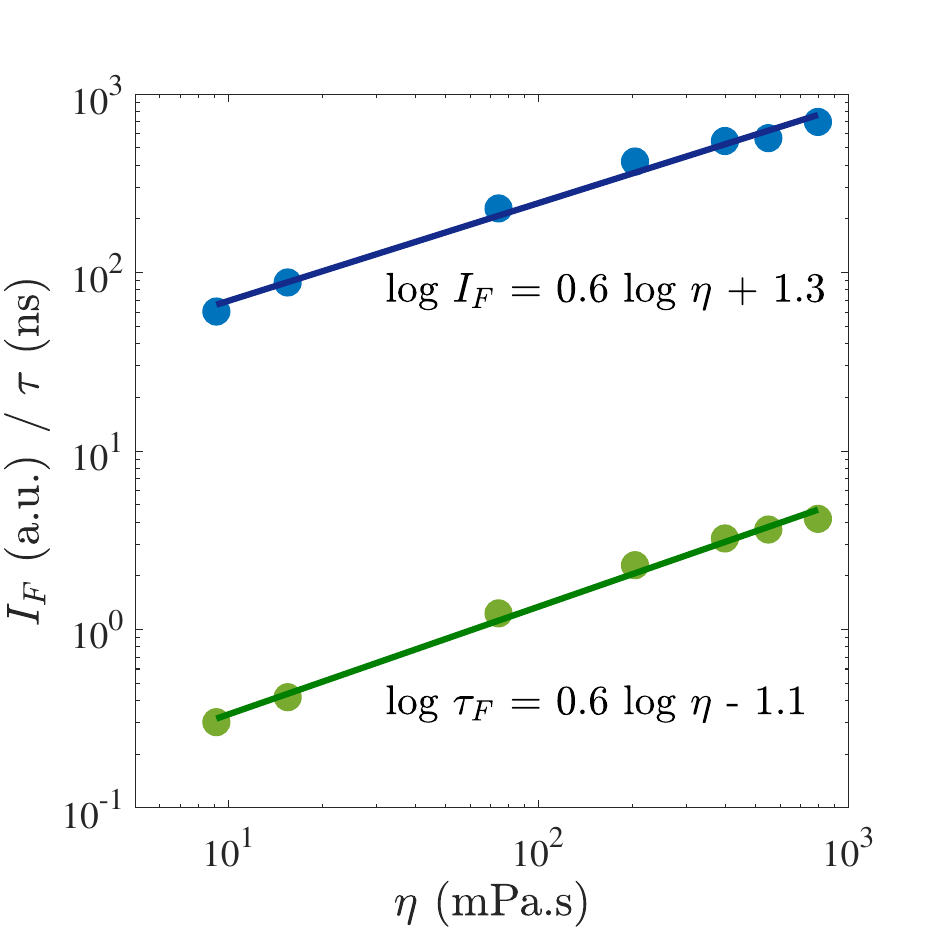}
\end{minipage}%
\caption{(a) Emission spectra of BODIPY-2-OH in different DMSO-glycerol mixtures of varying viscosities. The black vertical line  at $\SI{515}{\nano\meter}$ shows that the emission maximum wavelength is independent of the viscosity of the samples. (b) Calibration curves in logarithmic scale of BODIPY-2-OH fluorescence intensities, $I_\text{F}$ (blue points) and fluorescence lifetimes, $\tau_\text{F}$ (green points) versus DMSO-glycerol mixture viscosity, $\eta$. The straight lines correspond to fits with the Förster-Hoffmann model (Equation~\ref{FH equation}) with similar exponent $\alpha=0.6$ for both fluorescence lifetime and intensity measurements, and prefactor $C_I=1.3$ for fluorescence intensity and $C_\tau =-1.1$ for lifetime measurement when $\eta$ is expressed in $\SI{}{\milli\pascal\second}$ and $\tau$ in $\SI{}{\nano\second}$. Error bars for $I_\text{F}$ and $\tau_\text{F}$ are smaller than the size of data points.}
\label{fig:charac_BODIPY}
\end{figure*}

The emission spectra of solutions of BODIPY-2-OH in DMSO/glycerol mixtures of different concentrations, and thus of varying viscosity, are reported in Figure~\ref{fig:charac_BODIPY}(a). It is first interesting to note that no changes of absorption ($\lambda_\text{abs} = \SI{500}{\nano\meter}$) and emission ($\lambda_\text{abs} = \SI{515}{\nano\meter}$) maxima wavelengths were observed, which suggests that fluorescence response is not affected by polarity changes in the investigated DMSO/glycerol mixtures~\cite{Dent-Kuimova2015}.

The variations of fluorescence intensity $I_\text{F}$ (acquired with a fluorescence spectrometer) and lifetime $\tau_\text{F}$ (acquired with FLIM) with viscosity are displayed in Figure~\ref{fig:charac_BODIPY}(b). Both parameters increase with viscosity, which qualitatively agrees with the general mechanism of FMR. This confirms that BODIPY-2-OH can be used as a local viscosity probe. Lifetime values of the probe measured for $\eta > \SI{9}{\milli\pascal\second}$ are within the detection limit of the LIFA-FLIM used for this work. 

More quantitatively, both parameters evolve with viscosity as a power-law with similar exponent $\alpha=0.6$ over about three orders of magnitude in viscosity. Such an observation is consistent with previous measurements in the literature for other BODIPY-based FMR~\cite{Wu-Kuimova2013,Kuimova2013,Kuimova2012}. 

A power-law increase of fluorescence properties with ambient viscosity is usual with FMR and derives from Förster-Hoffmann equation~\cite{ForsterHoffmann1971}. Quantum yield, $\phi_\text{F}$ of FMR has indeed been proposed to follow a power-law relationship with viscosity along:
\begin{equation}
  \log \phi_{F} = \alpha \log {\eta}+ C
  \label{FH equation}
\end{equation}
\noindent where $\phi_{F}$ and $\eta$ represent the fluorescence quantum yield and the local viscosity respectively; whereas $\alpha$ is a dye-dependent constant and $C$ is a proportionality constant~\cite{Loutfy1986}. As the steady-state intensity and fluorescence lifetime are proportional to the quantum yield, they follow a similar Förster-Hoffmann relationship as expressed below\cite{JournalReviewHaidekker2010,Suhling2007} :
\begin{equation}
  \log I_{F} = \alpha \log {\eta}+ C_{I_{F}} 
  \label{eq:FH-Intensity}
\end{equation}
\noindent and:
\begin{equation}
 \log \tau_{F} = \alpha \log {\eta}+ C_{\tau_{F}} 
  \label{eq:FH-Lifetime}
\end{equation}
\noindent Both $I_\text{F}$ and $\tau_\text{F}$ are thus expected to follow a power-law relationship with viscosity, sharing a similar exponent $\alpha$ and possibly different pre-factors $C$, which is in agreement with experimental results.

\begin{figure}[!htb]
    \centering
    \includegraphics[scale=0.5]{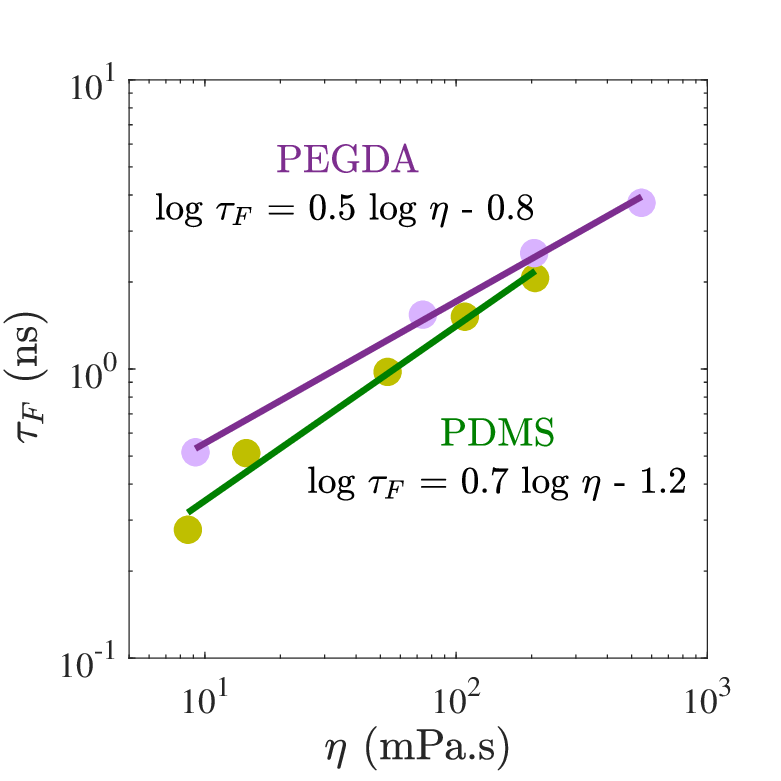}
    \caption{Logarithmic-scale calibration curve of BODIPY-2-OH fluorescence lifetimes, $\tau_\text{F}$ versus DMSO-glycerol mixture viscosity, $\eta$ in a PDMS microchannel (green dots) and in a PEGDA microchannel (purple dots). Straight lines correspond to a fit with the Förster-Hoffmann model (Equation~\ref{FH equation}) with exponent $\alpha$ = 0.7 for a PDMS microchannel and $\alpha$ = 0.5 for a PEGDA microchannel. Error bars for $\tau_\text{F}$ values are smaller than the size of data points.}
    \label{FH_chips}
\end{figure}

It is known that the environment can affect the fluorescence response of FMR \cite{Haidekker2007,JournalReviewHaidekker2010} and it is thus essential to perform calibrations \textit{in situ}. The calibration procedure for fluorescence lifetime was thus repeated directly within microfluidic channels, that will be used in the remaining of the paper. Channels were filled with various DMSO-glycerol mixtures of known viscosity containing BODIPY-2-OH, and the flow was left to stabilize for 1 minute before measurement. Different images were taken along the channel for each mixture and showed no significant lifetime variations. Average lifetime was then used to construct the calibration curves in Figure~\ref{FH_chips} for two different materials of the microfluidic chip.

Förster-Hoffmann relation~\eqref{eq:FH-Lifetime} thus remains valid for both materials, and measured lifetimes are of a similar order of magnitude to those measured in solution. However, there was a slight difference in the value of the exponent $\alpha$: while this does not compromise the use of the technique, it showed that \textit{in situ} calibrations are preferable in order to retrieve quantitative values of viscosity~\cite{Bunton_HeleShaw}. In the following, results of this \textit{in situ} calibration are used to obtain viscosity from measured lifetimes.

\subsection{Viscosity mapping of diffusive mixing in a simple Y-mixer}

In order to prove that FMR can be used for viscosity mapping in heterogeneous systems, experiments were performed in a simple Y-mixer, as depicted on Figure~\ref{fig:chip_schematics}(a). In this situation, two miscible DMSO/glycerol mixtures of different compositions (hence viscosity) were injected in the entrance sleeves of the microchannel. There, streams follow a laminar flow and mixing occurs only by transverse diffusion, perpendicular to the flow direction~\cite{Salmon2007}. This is a well-known configuration in microfluidics and the evolution of the fluid composition along the channel has been thoroughly studied. Studies of similar cases using FLIM can be found in literature but employed fluorescent probes whose lifetime evolved through quenching by a diffusing ion~\cite{Elder2006} or by changes of solvent polarity~\cite{Magennis2005}. Other measurements using molecular rotors have also been proposed before~\cite{Benninger2005} but exploited a more complex setup, characterizing decay of fluorescence anisotropy or measuring lifetime in time domain (thus requiring a pulsed laser source), and no quantitative comparison with models was presented. The most quantitative work that was found proposes to determine directly the composition through in situ Raman spectroscopy~\cite{Salmon2005,Dambrine2009,Salmon2007} and was taken as a reference for comparison with our work: a similar chip design was thus used. In particular, the channel aspect ratio was kept identical to ensure a unidimensional flow profile, and flow rates were adapted to have small Reynolds numbers. However, for practical reasons, narrower channels of a factor of about 2 were used.

For two immiscible fluids of different viscosity co-flowing along a channel, the fraction of the channel occupied by a given stream is proportional to the product of its viscosity and average flow rate~\cite{Yager2003}. For this experiment, the viscosities of the two incoming DMSO/glycerol mixtures were $\eta_{1} \approx \SI{9}{\milli\pascal\second}$ and $\eta_{2} \approx \SI{74}{\milli\pascal\second}$ with a ratio of about $1:8$. Experiments were conducted with flow rates adjusted to keep the average interdiffusion zone visible along the entire width of the channel.

\begin{figure*}[!htb]
\parbox{0.4\textwidth}{\includegraphics[scale=0.6]{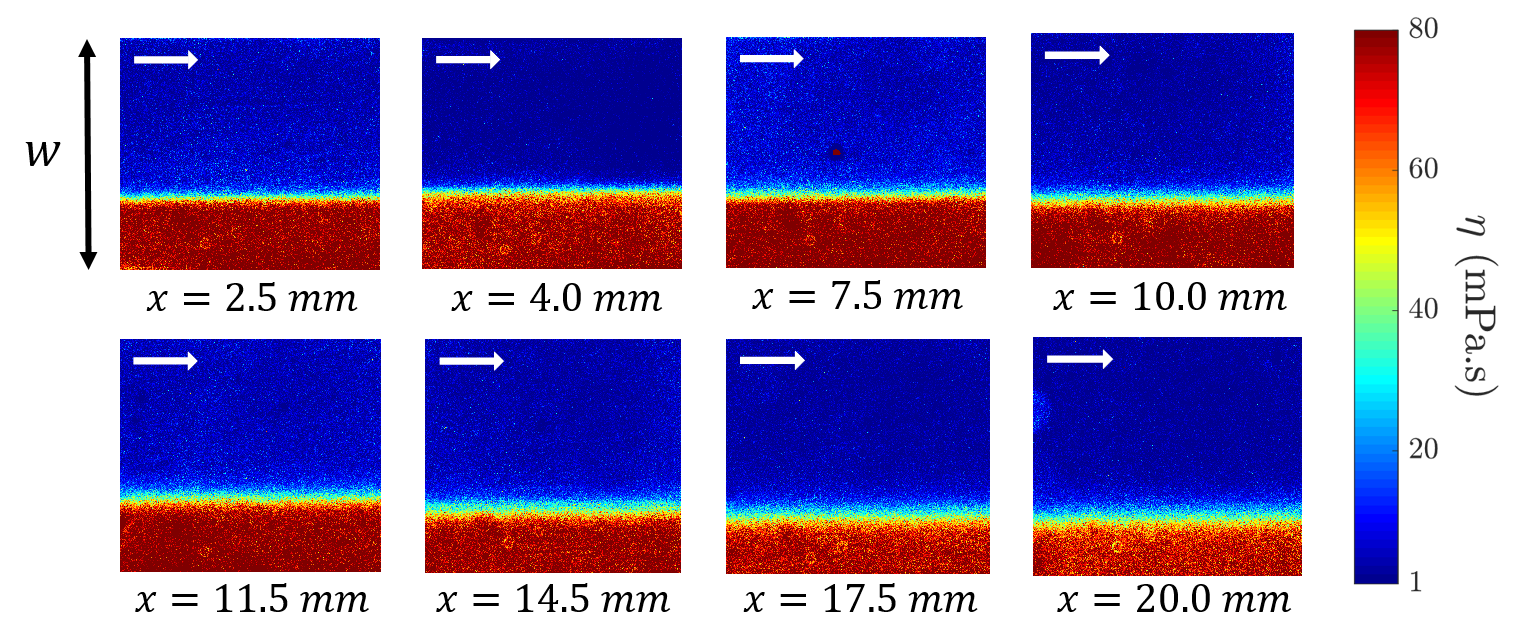}}%
\newline
\begin{minipage}{0.4\textwidth}%
\includegraphics[scale=0.6]{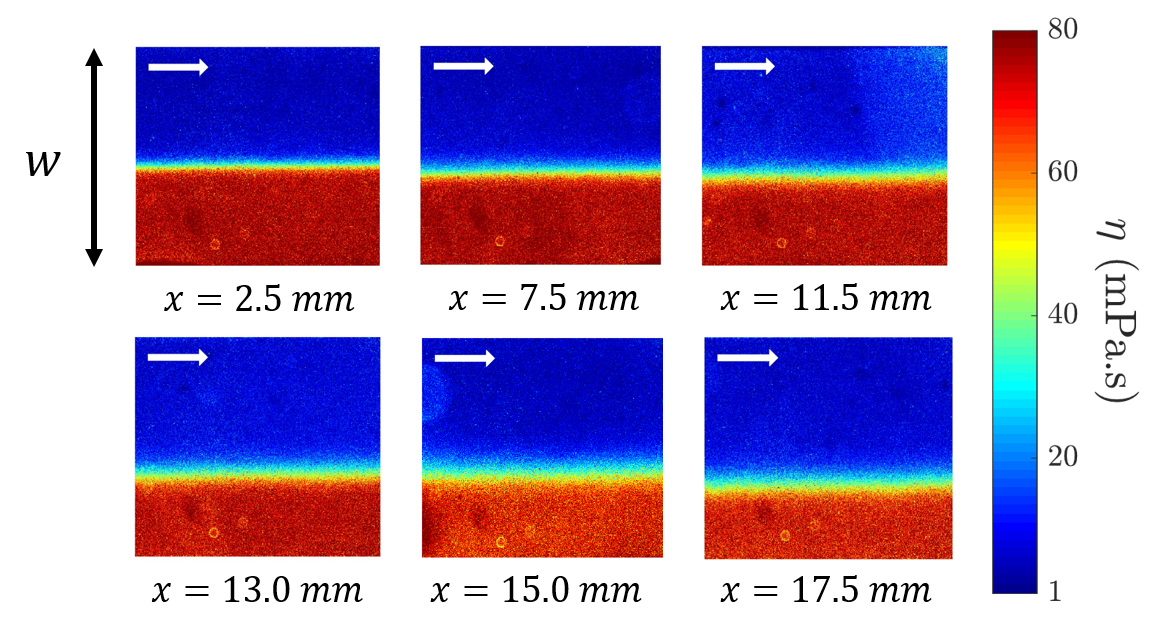}
\end{minipage}%
   \caption{Viscosity mapping during of two liquid streams (denoted by S1, upper stream, and S2, lower stream) flowing in a simple Y-mixer. Injected DMSO-glycerol mixtures have initial viscosity $\eta_1=\SI{9}{\milli\pascal\second}$ and $\eta_2=\SI{74}{\milli\pascal\second}$. Images are taken at different positions $x$ along the microchannel, $x=0$ corresponding to the first point of contact of the two liquids. The white arrow indicates the flow direction and corresponds to a length of $\SI{150}{\micro\meter}$. Flow rates of the two streams are respectively $(Q_\text{S1},Q_\text{S2}) =$ (a) $(11,1) \, \SI{}{\micro\liter\per\minute}$ and (b) $(7,1) \, \SI{}{\micro\liter\per\minute}$. \label{fig:visco_map_CF_all}}
\end{figure*}

FLIM technique allowed the measurement of local fluorescence lifetime at every pixel on different pictures taken along the microchannel at steady state. Using the calibration equation in Figure~\ref{FH_chips}, viscosity maps could be obtained as displayed in Figure~\ref{fig:visco_map_CF_all} for two different sets of flow rates. It is to note that a few outliers were observed on pictures, corresponding to inconsistent computed lifetimes far below the detection limit of LIFA ($< \SI{0.2}{\nano\second}$): these are, however, in limited number and were not taken into account for further data analysis.

The two streams of initially distinct viscosity progressively mix across an interdiffusion layer that widens along the channel. As can be seen by comparing Figure~\ref{fig:visco_map_CF_all}(a) and (b), a slower flow leads to better mixing at the end of the channel, as it corresponds to a longer time of contact between the solutions. Such a phenomenon will be further quantified in the discussion (see in particular Eqs.~\eqref{eq:timescale} and~\eqref{eq:diffusive_behaviour}). Finally, the interdiffusion layer is closer to the center of the channel for a ratio of flow rate close to the ratio of viscosity, as would be the case for immiscible liquids. A progressive drift of the interdiffusion zone towards the higher-viscosity fluid can also be observed. This latter fact results from the coupling between hydrodynamics and the mixing through the dependence of the viscosity with the glycerol concentration along the mixing channel~\cite{Dambrine2009}.

\subsection{Micromixing experiments}

Finally, the more complex situation of mixing of two liquid streams in a SHM channel was studied, following a similar protocol. The design of asymmetrical herringbone grooves that was used helped to develop laminar chaotic flows in the microchannel by stretching and folding the two incoming steady streams, into alternate thin liquid sheets in order to enhance diffusion efficiency~\cite{Stroock2002, Stone2004}. In order to observe the evolution of the mixing along the channel, zones with SHM were separated by zones of free diffusion, where flow returns to its laminar state and the degree of mixing can be measured.

In this experiment, the mixing of streams of initial viscosity $\eta_{1} = \SI{9}{\milli\pascal\second}$ and $\eta_{2} = \SI{206}{\milli\pascal\second}$ in such a system was studied. The lifetimes measured by LIFA-FLIM were converted into viscosity by using the calibration equation for the PDMS microchannel displayed in Figure~\ref{FH_chips}. 

\begin{figure*}[!htb]
\centering
\parbox{0.43\textwidth}{\includegraphics[width=0.43\textwidth]{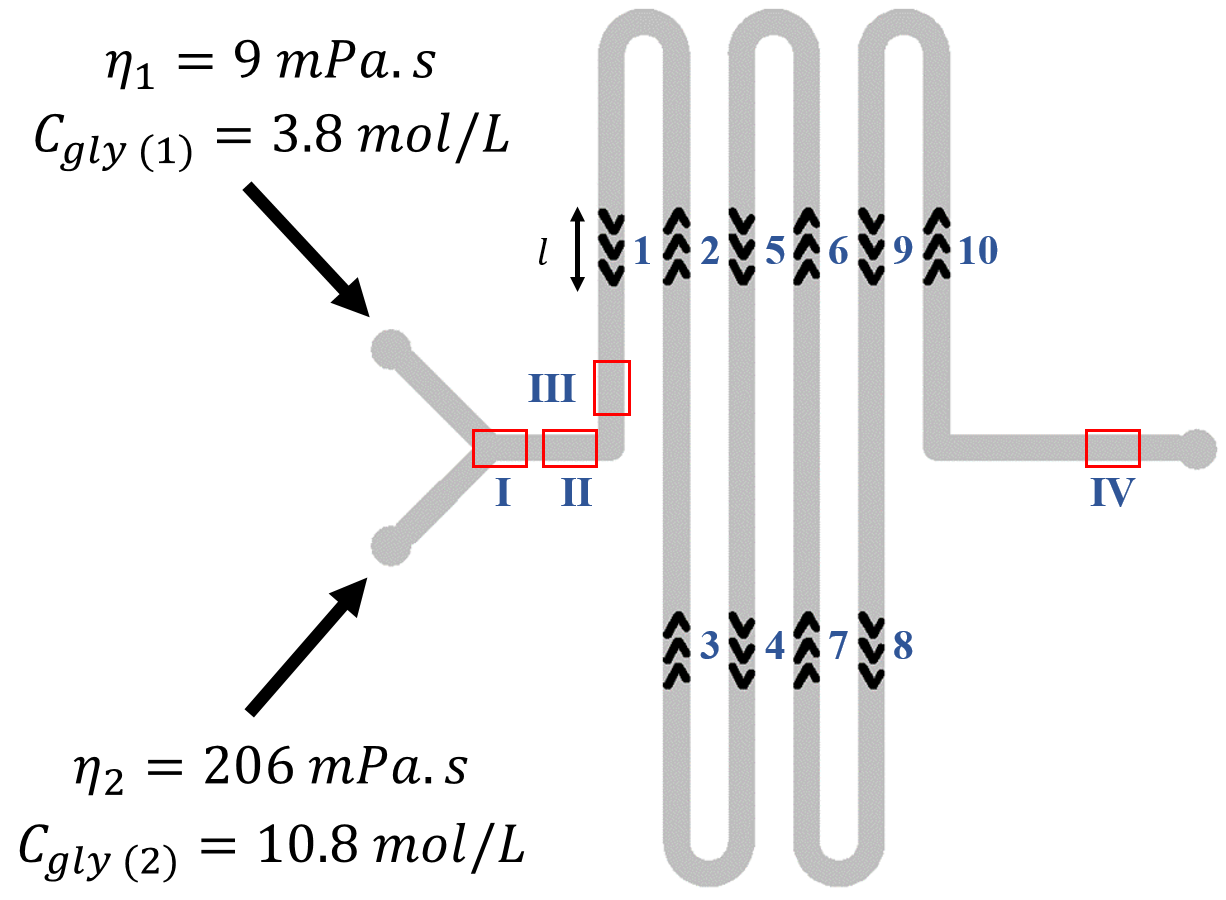}}%
\newline
\begin{minipage}{\textwidth}%
\includegraphics[width=\textwidth]{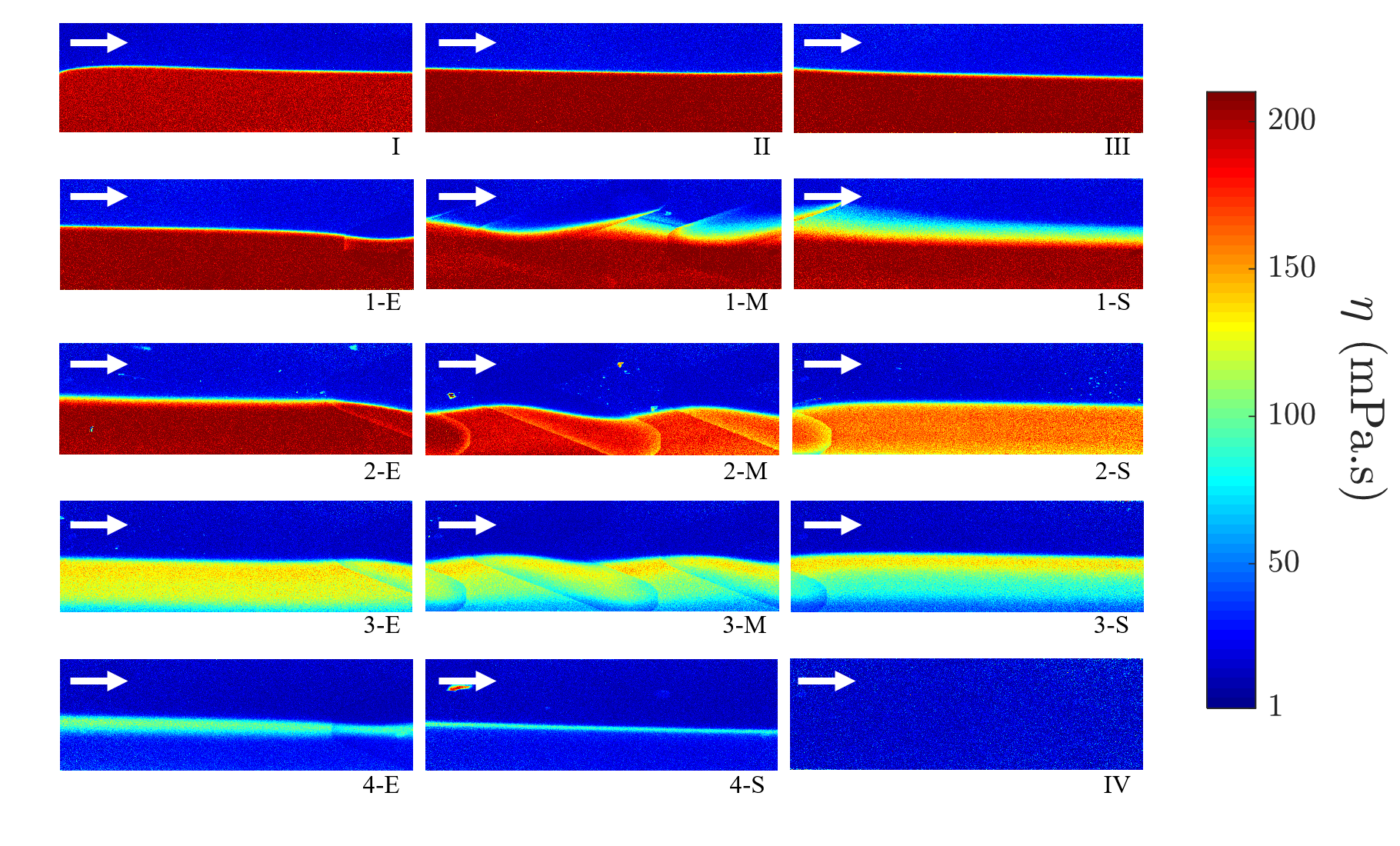}
\end{minipage}%
\caption{(Top) Schematics of a microfluidic Y-mixer with staggered herringbone passive micromixers (SHM). (Bottom) Viscosity mapping in a Y-mixer with SHM during a co-flow of DMSO-glycerol mixture of initial viscosity $\eta_1=\SI{9}{\milli\pascal\second}$ (S1) and $\eta_2 = \SI{206}{\milli\pascal\second}$ (S2). Applied flow rates are $(Q_\text{S1},Q_\text{S2})=(35,0.25) \, \SI{}{\micro\liter\per\minute}$. The arrow represents the flow direction and a scale length of $\SI{200}{\micro\meter}$. Images were taken at different positions along the length of the microchannel, labeled on the top schematics. Indications E, M, and S, respectively stand for inlet, middle zone, and outlet.}  
\label{visco_map_MM}
\end{figure*}

The obtained viscosity maps at different positions along the channel are depicted in Figure~\ref{visco_map_MM}. The different positions are represented on the schematics on top of the figure: in particular, pictures designated by E correspond to the entrance of SHM zones, where the flow is laminar. In these pictures, after every SHM, the mixing degree increased, which was characterized by the widening interdiffusion layer and the homogenization of the viscosities of the two streams. Pictures designated by M and S are taken in and after SHM zones, respectively, where we can observe a destabilization of the interdiffusion layer, leading to enhanced mixing.

\section{Discussion}
\label{sect:discussion}

In this last part, viscosity mappings presented in the Results section are analyzed in more detail, confirming the relevance of FMR and of the FLIM setup for quantitative measurements.

\subsection{Effective diffusion coefficient in the simple Y-mixer}

In order to assess the validity of the viscosity measurements, it is possible to analyze viscosity maps of the simple co-flow, displayed in Fig.~\ref{fig:visco_map_CF_all} in further details. In this situation, it is indeed possible to compare the results with a model from the literature.~\cite{Salmon2005,Dambrine2009,Salmon2007} 

\begin{figure}[!htb]
    \centering
    \includegraphics[scale=0.5]{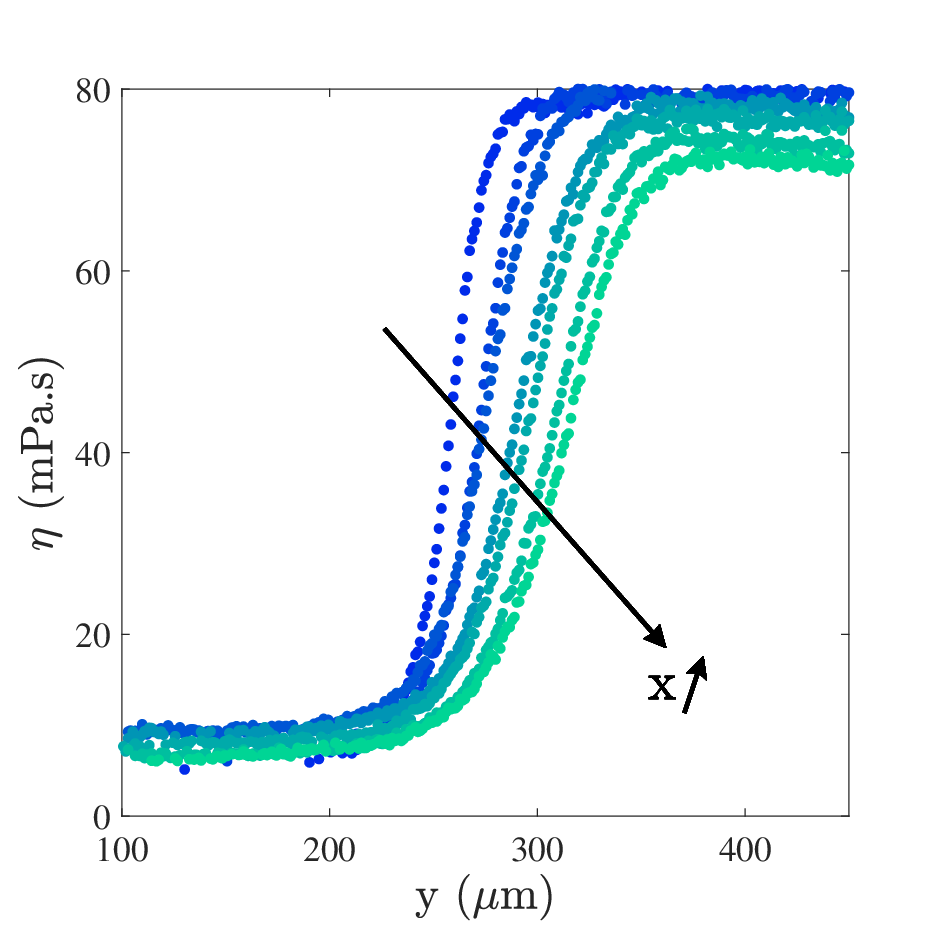}
    \caption{Viscosity profiles $\eta$ versus $y$, for different positions $x$ along the microchannel (color-coded: darker to lighter shades represent increasing $x$ direction, downstream), in a simple Y-mixer with flow rates $(Q_\text{S1},Q_\text{S2}) = (7,1) \, \SI{}{\micro\liter\per\min}$. The arrow represents downstream evolution, with growing $x$.}
    \label{Visco_Profile_CF1}
\end{figure}

From experimental viscosity maps at different positions $x$ in the channel, transverse viscosity profiles $\eta(y)$ can first be extracted, with $y \in [0,w]$ corresponding to the coordinate perpendicular to the flow direction. As diffusion is slow enough, profiles on a single picture do not significantly evolve with $x$: profiles were thus averaged in all pictures to minimize noise level. The results are displayed in Figure~\ref{Visco_Profile_CF1}.

With these profiles, the evolution of viscosity across the channel can be observed more clearly, and in particular the widening of the interdiffusion layer along the channel. This situation of diffusive mixing in a microfluidic co-flow can be modeled by an advection-diffusion problem. In the case of an infinitely wide channel, analytical solutions have been obtained for the volume fraction $\phi$ of glycerol~\cite{Salmon2007}:
\begin{equation}
\centering
  {\phi}\left(x,y\right) = \frac{\phi_{0}}{2} \erf \left(\frac{y-y_m(x)}{2\sigma_d (x)}  \right) + \phi_\text{m}.
  \label{Deff_fitting_eq}
\end{equation}
\noindent At a given position $x$ in the channel, $\phi$ varies from $\phi_\text{min} = \phi_\text{m} - \phi_0/2$ to $\phi_\text{max} = \phi_\text{m} + \phi_0/2$, with a transition described by an error function, of center $y_m(x)$ and characteristic width $\sigma_d(x)$. 

Such an expression has been experimentally validated~\cite{Dambrine2009} and was used as a reference in our work. In particular, flow rates were selected to maintain a low Reynolds number, typically between $\SI{e-2}{}$ and $\SI{e-1}{}$, in order to obtain laminar flow in the microchannels.  Also, the entrance length $L_\text{e}$, of about $\SI{0.4}{\milli\meter}$, was small enough to consider that the flow reached a fully developed Poiseuille profile in the analyzed pictures~\cite{Yager2003, Stiles2004}.

\begin{figure*}[!htb]
\centering
\parbox{0.45\textwidth}{\includegraphics[width=0.45\textwidth]{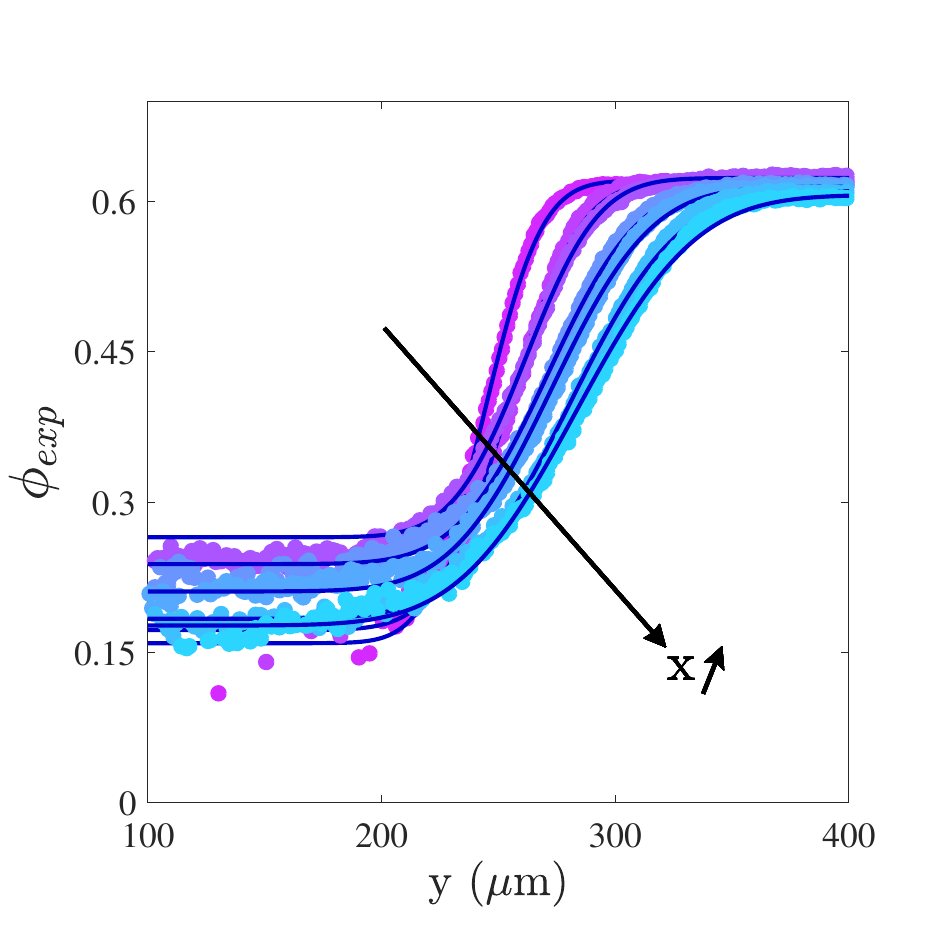}}%
\begin{minipage}{0.45\textwidth}%
\includegraphics[width=\textwidth]{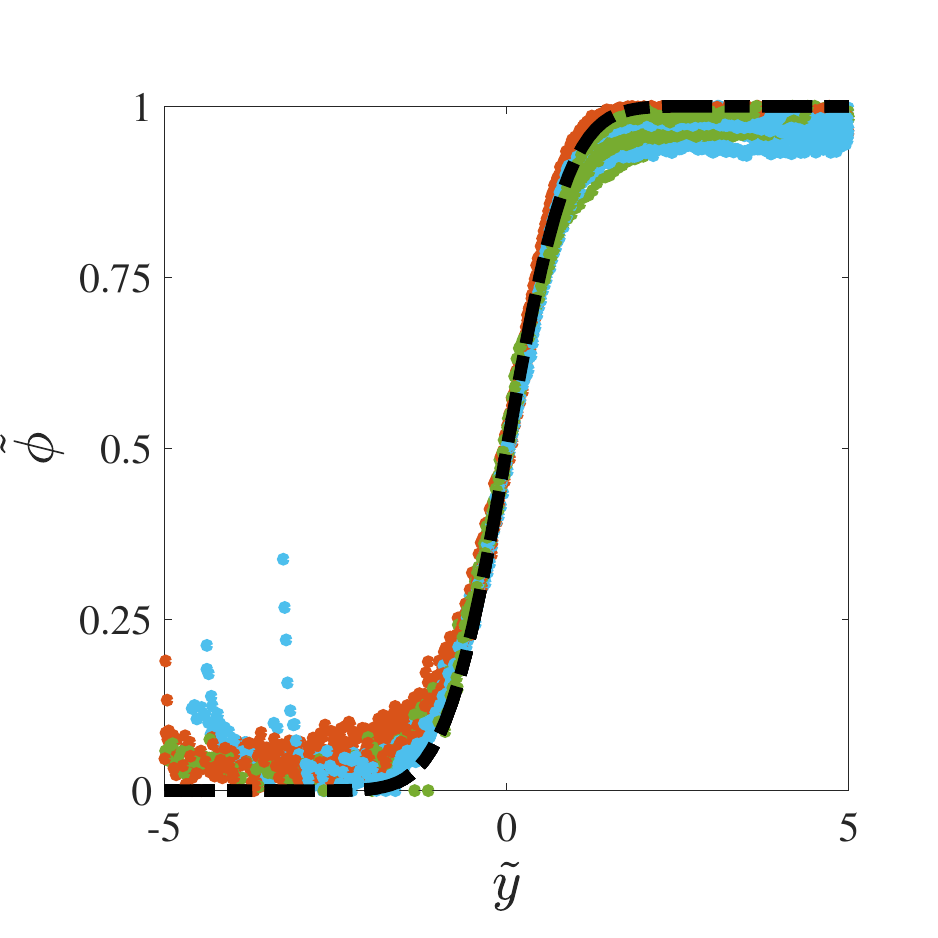}
\end{minipage}%
\caption{(a) Profiles of glycerol volume fraction ${\phi_\text{exp}}(y)$ at different positions $x$ along the micro-channel (color-coded: darker to lighter shades represent increasing $x$ direction, downstream) obtained from viscosity profiles obtained in Figure~\ref{Visco_Profile_CF1}. The continuous lines represent the fits according to Equation~\ref{Deff_fitting_eq}. (b) Normalized glycerol volume fraction $\tilde{\phi}$ versus normalized position $\tilde{y}$ as defined in Equation~\eqref{eq:normalized_phi}. Data obtained for different flow rates $(Q_\text{S1},Q_\text{S2})=(7,1)$ (brown), $(3,1)$ (blue) and $(11,1) \, \SI{}{\micro\liter\per\min}$ (blue). The black dashed line represents the theoretical master curve from  Equation~\eqref{eq:mastercurve}.}
\label{fig:Fit_VolFrac_CF1}
\end{figure*}

In order to use Equation~\eqref{Deff_fitting_eq}, the measured viscosity $\eta$ was first converted into glycerol volume fraction $\phi$ by using a pre-established calibration curve (see Supporting Information): corresponding profiles at different positions $x$ are displayed in Figure~\ref{fig:Fit_VolFrac_CF1} and were fitted using Equation~\eqref{Deff_fitting_eq}. As also observable on the viscosity profile, the minimum and maximum volume fractions at both sides are not constant along the microchannel. This was not observed in previous experiments from the literature and comes from the smaller width $w$ of the channel, which induces side effects not considered in the model and become observable in our experiment. Parameters $\phi_0$, $\phi_m$, $\sigma_d$, and $y_m$ were thus taken as free fit parameters at every position $x$. The resulting fits are superimposed to experimental data on Figure~\ref{fig:Fit_VolFrac_CF1}.

First, in order to assess the accuracy of the proposed model, the normalized volume fraction:
\begin{equation}
    \tilde{\phi} = \frac{\phi - \phi_\text{min}}{\phi_\text{max} - \phi_\text{min}}
    \label{eq:normalized_phi}
\end{equation}
\noindent can be plotted as a function of a normalized coordinate normal to the channel $\tilde{y} = (y-y_m(x))/2\sigma_d(x)$. According to the model~\eqref{Deff_fitting_eq}, this should collapse all data on a single master curve described by:
\begin{equation}
    \tilde{\phi}(\tilde{y}) = \frac{1 + \erf \tilde{y}}{2}
    \label{eq:mastercurve}
\end{equation}
\noindent This is indeed observed in Figure~\ref{fig:Fit_VolFrac_CF1}(b), where datasets corresponding to different values of the flow rates $Q_{S1}$ and $Q_{S2}$ are represented. This confirms the agreement of the model with the measurements.

The variations of the fitting parameters with the position $x$ along the microchannel can be characterized in more details. As commonly done in microfluidics, the average time $\tau$ spent by the fluid in the channel at a given position $x$ is used to describe the evolution of the system:
\begin{equation}
\centering
  \tau= \frac{2xhw}{Q_{S1}+Q_{S2}} = \frac{2x}{v_1+v_2}    
  \label{eq:timescale}
\end{equation} 
\noindent with $v_1$ and $v_2$ representing the velocities of the incoming streams.

\begin{figure*}[!htb]
\centering
\parbox{0.3\textwidth}{\includegraphics[width=0.3\textwidth]{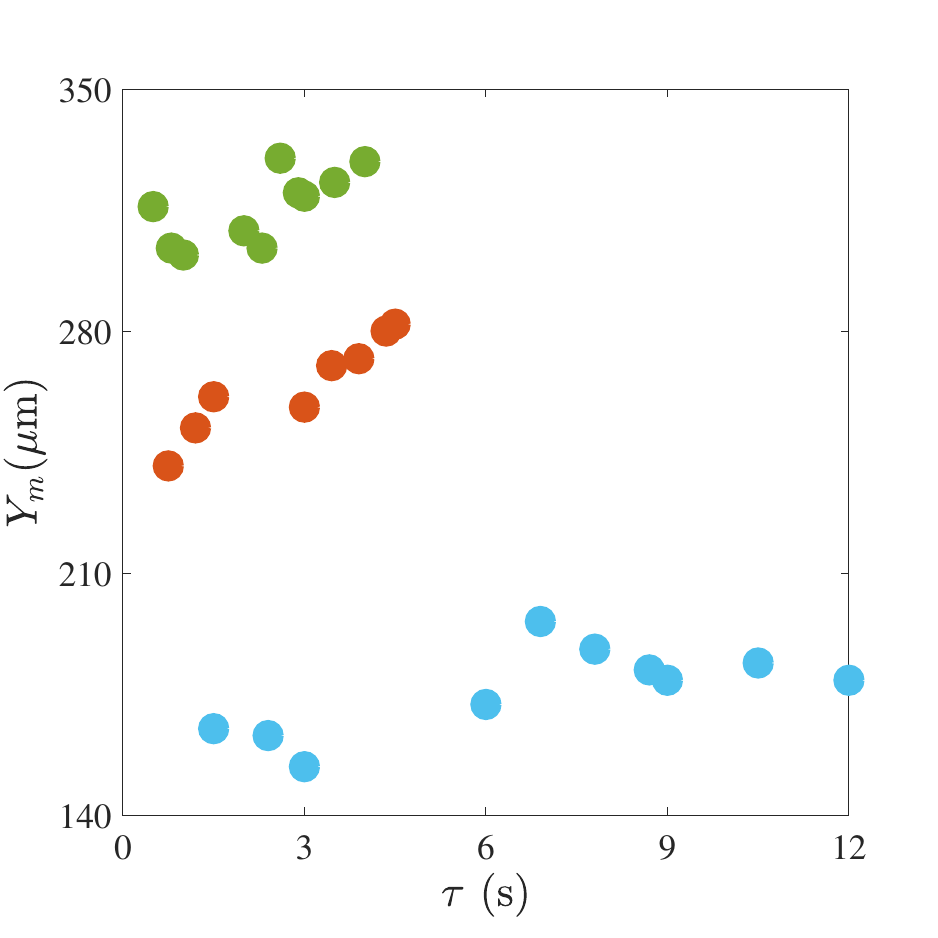}}%
\begin{minipage}{0.3\textwidth}%
\includegraphics[width=\textwidth]{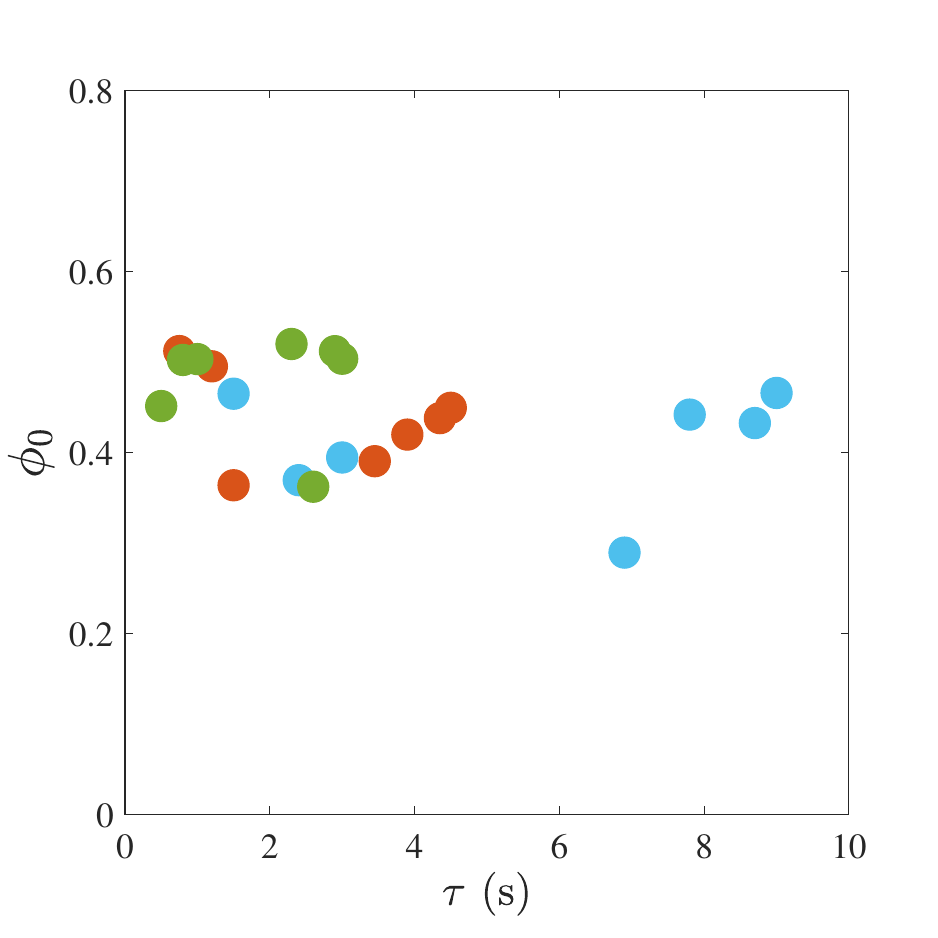}
\end{minipage}%
\begin{minipage}{0.3\textwidth}%
\includegraphics[width=\textwidth]{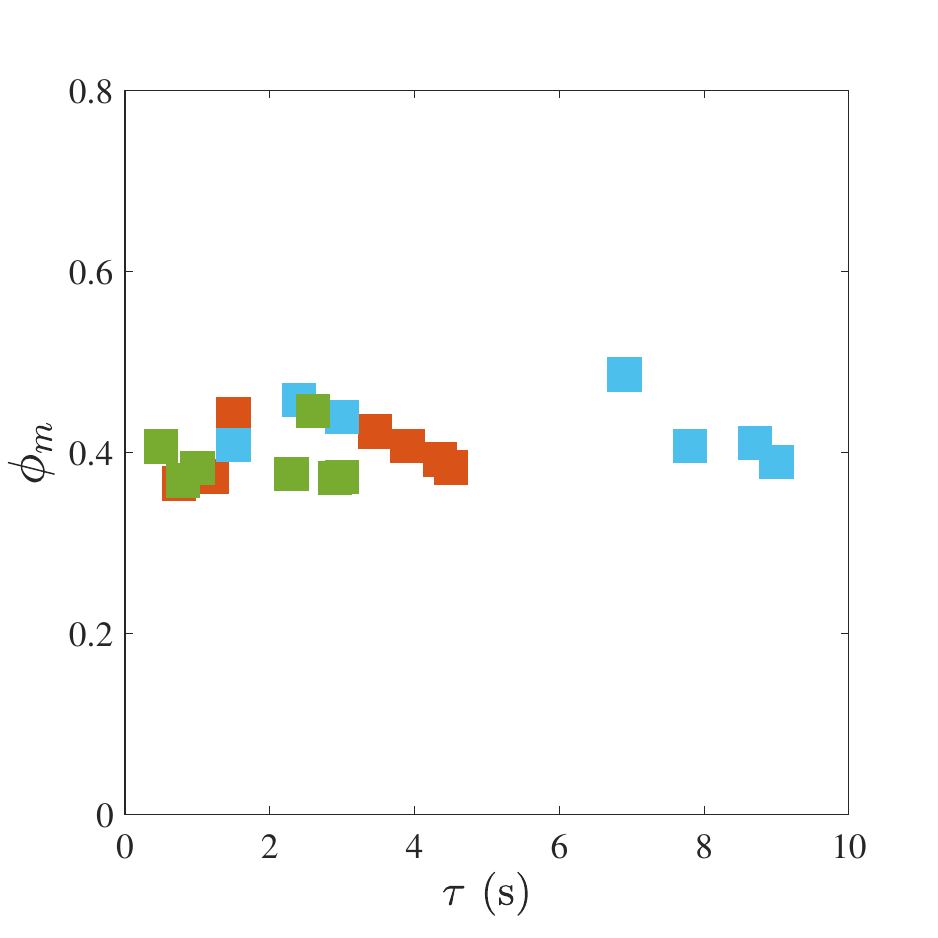}
\end{minipage}%
\caption{Variation of fitting parameters (a) $y_m$, (b) ${\phi_0}$, and (c) ${\phi_{min}}$ versus characteristic time $\tau$ obtained during the co-flow experiments in a simple Y-mixer with ($Q_\text{S1},~Q_\text{S2}$) = $(7,1)$ (brown); $(3, 1)$ (blue); $(11,1) \,\SI{}{\micro\liter\per\min}$ (green).}
\label{fig:fit_constant}
\end{figure*}

The evolution of the amplitude $\phi_0$ of the profile and the volume fraction at the middle of the interdiffusion layer $\phi_m$ are given in Figure~\ref{fig:fit_constant}(b) and Figure~\ref{fig:fit_constant}(c). These parameters are difficult to interpret, but they are found to be roughly constant within a $\SI{10}{\percent}$ variation (more precisely $\phi_0 = \SI{44(10)}{\percent}$ and $\phi_m = \SI{40(6)}{\percent}$). This may seem contradictory with the clear evolution of the viscosity on the sides of the channel: it is however easily explained by the strongly non-linear evolution of viscosity of glycerol-DMSO mixture with glycerol concentration. Due to the finite width of the channel, the volume fractions at the walls of the channel slightly evolve, which induces a strong viscosity variation for the stream highly concentrated in glycerol.

The position of the center of the interdiffusion layer $y_m$ is also given in Figure~\ref{fig:fit_constant}(a). Again, this parameter is roughly constant, showing a tendency to increase, which becomes clearer when the flow rate ratio differs from the viscosity ratio of the two streams. This is consistent with observations by Dambrine et al.~\cite{Dambrine2009}.

\begin{figure}[!htb]
    \centering
    \includegraphics[scale=0.5]{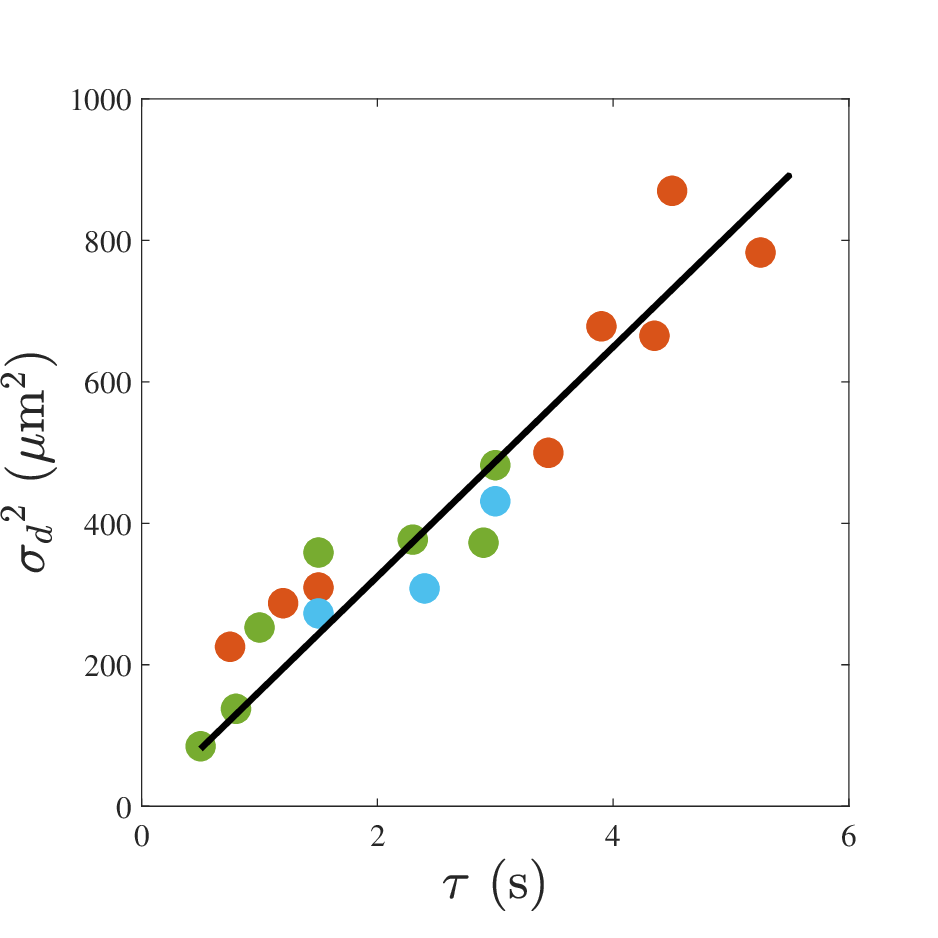}
    \caption{Evolution of the squared interdiffusion layer width, $\sigma_d^2$, against the timescale, $\tau$, defined in Equation~\eqref{eq:timescale}, with ($Q_\text{S1},~Q_\text{S2}$) = $(7,1)$ (brown); $(3, 1)$ (blue); $(11,1) \,\SI{}{\micro\liter\per\min}$ (green). Continuous line illustrates fit by Equation~\eqref{eq:diffusive_behaviour} with effective diffusion coefficient (averaged for all flow rates), $D_\text{eff} = \SI{2e-10}{\meter\squared\per\second}$.}
    \label{Fig_Deff}
\end{figure}

More interestingly, the evolution of width $\sigma_d$ of the interdiffusion layer, which is displayed in Figure~\ref{Fig_Deff}, is predicted by the model. As the spreading of this layer is driven by transverse diffusion, the squared width $\sigma^2_d(x)$ at a given position $x$ is expected to evolve linearly with the time of diffusion $\tau(x)$ along:
\begin{equation}
    \sigma_d^2(x) = D_\text{eff} \, \tau(x).
  \label{eq:diffusive_behaviour}
\end{equation}
\noindent Experimental observations are in agreement with this relation, and allow to retrieve an effective diffusion coefficient $D_\text{eff} = \SI{2e-10}{\meter\squared\per\second}$, to be compared with the reference value $D_\text{ref} = \SI{7e-10}{\meter\squared\per\second}$ from the literature~\cite{Dambrine2009}. These two values have a similar order of magnitude, and this agreement can be considered as satisfactory. First, the reference value obtained from literature was for water-glycerol mixtures while DMSO-glycerol mixtures were used in this study. Also, no systematic determination of experimental uncertainties was performed, but considering the number of steps to retrieve this diffusion coefficient, uncertainties are likely to be significant. Finally, the existence of a finite size effect due to the narrower channels employed in this study can also slightly modify the effective diffusion coefficient.

As a conclusion, these results show that the technique allows mapping of viscosity profile in a Y-mixer configuration which agrees with experiments and models proposed in the literature. This validated the use of such setup for quantitative measurements of viscosity in microfluidic chips, and more generally in confined flows.

\subsection{Mixing in passive micromixers}

While the flow is more complex in micromixers than in the simple Y-mixer, the viscosity maps displayed in Fig.~\ref{visco_map_MM} can also be analyzed to retrieve quantitative information on mixing.

\subsubsection{Final homogeneous state}

After crossing a few SHM, the two miscible streams of different viscosities eventually mix, leading to a homogeneous mixture. Using the principle of mass conservation, the final glycerol concentration of the homogeneous mixture $\mathrm{[Gly_{calc}]}$ was related to the flow rates $Q_\text{Si}$ and concentrations $\mathrm{[Gly_{Si}]}$ of the two incoming streams through:
\begin{equation}
    \mathrm{[Gly_{calc}]} = \frac{\mathrm{[Gly_{S1}]}.Q_\text{S1} + \mathrm{[Gly_{S2}]}.Q_\text{S2}}{Q_\text{S1} + Q_\text{S2}}.
    \label{CdM}    
\end{equation}
\noindent The final concentration of glycerol $\mathrm{[Gly_{exp}]}$ was measured by converting the measured viscosity into glycerol concentration with a pre-established viscosity-concentration curve in Supporting Information). Results are gathered in Table~\ref{table1} for different sets of incoming flow rates. The measured and calculated values are in good agreement, as can be seen, more quantitatively through their relative discrepancy $\sigma_\text{rel}$ remaining below $\SI{10}{\percent}$; this again validates the quantitative nature of the measurement method.

\begin{table*}[!htb]
\centering
\begin{tabular}{|c|c|c|c|c|}
 \hline
 $Q_\text{S1}$ ($\SI{}{\micro\liter\per\minute)}$ & $Q_\text{S2}$ ($\SI{}{\micro\liter\per\minute)}$ & $\mathrm{[Gly_{calc}]}$ ($\SI{}{\mol\per\liter}$) & $\mathrm{[Gly_{exp}]}$ ($\SI{}{\mol\per\liter}$) & $\sigma_\text{rel} (\%)$\\ 
 \hline
 7.0 & 0.5 & 4.28 & 4.21 & 1.5 \\ 
 \hline
 16.0 & 0.5 & 4.02 & 4.10 & 1.9 \\
 \hline
 4.0 & 0.5 & 4.59 & 4.19 & 8.6 \\
 \hline
\end{tabular}
\caption{Comparison of calculated glycerol concentration $\mathrm{[Gly_{calc}]}$ and experimental values $\mathrm{[Gly_{exp}]}$ after full mixing of the two streams in a SHM, for different flow rates.} 
\label{table1}
\end{table*}

\subsubsection{Assessment of mixing efficiency}
\label{subsec:4_1_micromixer_efficiency}

The obtained viscosity maps could be exploited more quantitatively to assess the efficiency of the micromixer. It is important to note that the objective of this study was not to provide an efficient micromixer design, but rather to propose and test a methodology that can be used to test the efficiency of micromixers while viscosity is mapped during mixing. As already mentioned, the flow in the SHM parts of the channel is three-dimensional and difficult to analyze: it is thus more relevant to focus on images taken before each of the SHM groups, where flow returned to its laminar state. In the following discussion, the mixer design was kept similar and the effect of injection flow rates is discussed.

In order to characterize the mixing efficiency, following previous approaches in the literature~\cite{Stroock2002, Enders2019}, the standard deviation of lifetime $\sigma$ was measured over the different pictures. In an ideal situation, $\sigma$ should be maximal at the entrance of the microchannel (and associated with a stepwise profile of lifetime), and decay to zero when the two fluids are homogeneously mixed. In practice, due to experimental noise, $\sigma$ reaches a plateau value at the end of the chip. In order to better compare the situations obtained with different flow rates, it is convenient to study the normalized parameter $\tilde{\sigma}$ defined by:
\begin{equation}
\centering
  \tilde{\sigma}=\frac{\sigma - \sigma_\text{min}}{\sigma_\text{max}- \sigma_\text{min}}         
  \label{eq:std_norm}
\end{equation}
\noindent where $\sigma_\text{min}$ is the minimum standard deviation measured at the outlet of the chip, and $\sigma_\text{max}$ is its maximum value measured at the entrance of the chip. Values of these extrema were similar within a $\SI{10}{\percent}$ variation for the different flow rates considered here. 

\begin{figure}[!htb]
    \centering
    \includegraphics[scale=0.45]{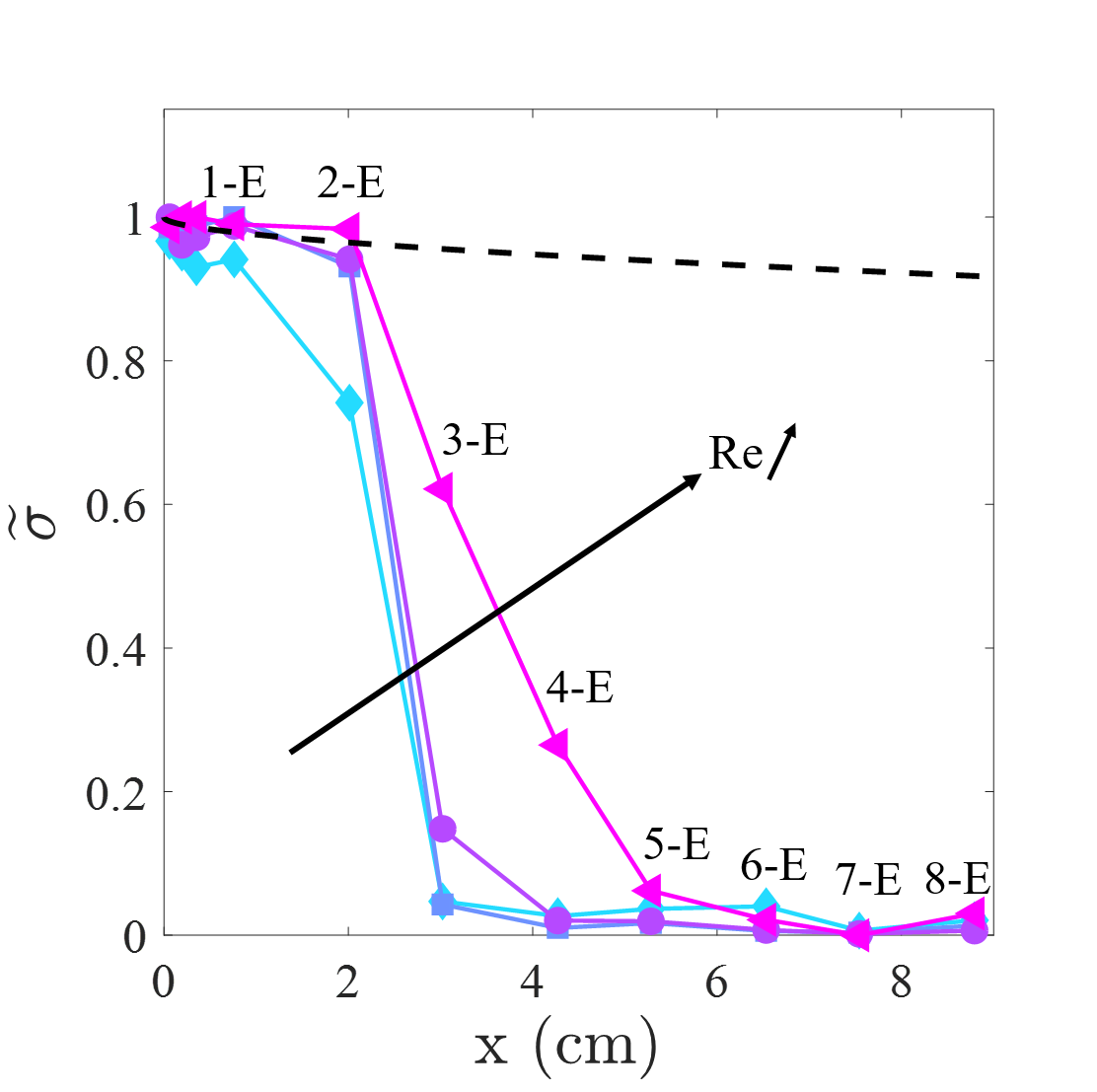}
        \caption{Normalized  standard deviation, $\tilde{\sigma}$, of the lifetime maps such as in Fig~\ref{visco_map_MM} defined in Equation~\eqref{eq:std_norm} as a function of the distance downstream from the entrance of the channel, $x$, at the beginning of every SHM group. Data shown in this figure were generated from a single microfluidic device during the mixing of DMSO-glycerol solutions of different viscosity, $\eta_1=\SI{9}{\milli\pascal\second}$ (S1) and $\eta_2=\SI{206}{\milli\pascal\second}$ (S2).
    Applied flow rates $(Q_\text{S1}, Q_\text{S2})$ were $(3.5, 0.25)$ ($\blacklozenge$, average Reynolds number $\mathrm{Re}=\SI{2e-2}{}$); $(7.0, 0.5)$ ($\blacksquare$, average Reynolds number $\mathrm{Re}=\SI{5e-2}{}$); $(14.0,1.0)$ ($\bullet$, average Reynolds number $\mathrm{Re}=\SI{0.1}{}$); $(35.0,2.5) \,\SI{}{\micro\liter\per\minute}$ ($\RHD$, average Reynolds number $\mathrm{Re}=\SI{0.2}{}$). Colored continuous lines are simply added for visual clarity only. The black dashed line represents the evolution that would be obtained for purely diffusive mixing in a simple Y-mixer.}
    \label{EfficiencyFigure}
\end{figure}

The obtained evolution of $\tilde{\sigma}$ after the different groups of SHM is represented in Figure~\ref{EfficiencyFigure}. For all studied flow rates, the mixing remained minimal before the entry of the first group of SHM (MM1), as also observed in Figure~\ref{visco_map_MM}. Then, mixing significantly improved after every group of SHM, and a homogeneous state is eventually reached.

The dashed line represents the evolution that would be observed without micromixers. Our results clearly illustrate the efficiency of micromixers compared to free diffusion. In SHM, the flow becomes three-dimensional and decomposes the streams into alternated thin sheets, in which diffusion becomes more efficient~\cite{Dreher2009,Enders2019}.

As also demonstrated in Figure~\ref{EfficiencyFigure}, stronger flow rates increased mixing efficiency in SHM, in agreement with previous studies~\cite{Stroock2002}. Care was taken to let the flow reach a steady state prior to any measurement after changing the flow rate. This, again, was in qualitative agreement with our understanding of micromixers: faster flows decrease contact time between the sheets in SHM zones, thus decreasing mixing efficiency.

The proposed method thus allows the assessment of the mixing performance of SHM micromixers, in agreement with results from the literature. This validates the use of molecular rotors to characterize viscosity in complex microfluidic flows.

\section{Conclusions}
\label{sec:conclusion}

In this work, the possibility to map quantitatively local viscosity in microfluidic flows using fluorescent molecular rotors was investigated. A BODIPY-based molecular rotor was synthesized, and its fluorescence lifetime response to viscosity was calibrated. In the well-controlled situation of the transverse diffusive mixing of two liquid streams flowing in a Y-mixer channel, the obtained viscosity maps were in quantitative agreement with models and experiments previously proposed in the literature for a similar system. This validates the possibility of using FMR for quantitative measurements beyond their previously reported use as contrast agents in bioimaging. A more complex system was then considered such as mixing in passive micromixers; it was then proved that FMR could be used to characterize mixing efficiency. It is important to note that they allow for measurement of up to three decades in viscosity, nearly up to \SI{1}{\pascal\second}.

This article proves the wide opportunities offered by FMR for quantitative and local characterization of viscosity in confined flows. In particular, the technique can be adapted to other solvents and tuned for a specific viscosity range by synthesizing new FMR. This opens a path for the characterization of fluids in numerous contexts, ranging from industrial processes to natural flows. It is however necessary to note the importance of performing \textit{in situ}  calibrations for accurate viscosity measurements, as the Förster-Hoffmann coefficients can be affected by the local environment of FMR.

In this work, all considered fluids were Newtonian and confined at the microscale. Future work will consider the characterization of complex fluids and of highly confined flows, close to the molecular scale. It will help to determine the spatial scale over which the viscosity of the microenvironment influences the response of FMR.

\textbf{Acknowledgements} \par 

The authors thank Dr. J.B Salmon and Dr. J. Leng for scientific exchanges concerning the fluorescence mapping in microfluidic chips. Dr. S. Harrisson is thanked for fruitful discussion. The research presented in this article was funded by ANR grant MicroVISCOTOR (ANR-18-CE42-0010-01). The authors also thank Solvay and CNRS for funding. This work was also supported by CONACYT (A1-S-7642), PAPIIT IN200422.

This document is the unedited Author's version of a Submitted Work that was subsequently accepted for publication in ACS Industrial Engineering and Chemistry Research, copyright 2023 American Chemical Society after peer review. To access the final edited and published work see \url{https://pubs.acs.org/doi/10.1021/acs.iecr.3c01047}.

\appendix

\section{Synthesis of molecular rotor}

Synthesis of the molecular rotor used in the paper follows the scheme represented in Figure~\ref{scheme1} of the main text, in the Material and Methods Section.

\subsection{Synthesis of Compound 1}

Compound (1), 4-(di(1H-pyrrol-2-yl)methyl)phenol was synthesized following a procedure in the literature~\cite{refMexico}. \textit{p}-hydroxybenzaldehyde ($\SI{1}{\gram}$, $\SI{8.19}{\milli\mole}$) and pyrrole ($\SI{5.49}{\gram}$, $\SI{81.88}{\milli\mole}$) were dissolved in anhydrous THF ($\SI{30}{\milli\liter}$) under $\mathrm{N_2}$ atmosphere followed by the addition of trifluoroacetic acid ($\SI{94}{\micro\liter}$, $\SI{1.23}{\milli\mole}$). The reaction mixture was stirred at room temperature for 15 minutes. The progress of the reaction was monitored by TLC. After $\mathrm{CH_2Cl_2}$ and 0.1 M $\mathrm{NaOH}$ were added to the reaction mixture and the organic phase was washed with water and filtered over $\mathrm{Na_2SO_4}$. Solvent and pyrrole were removed under reduced pressure. The crude product was purified by column chromatography over silica gel with hexane/ethyl acetate (7:3) as eluent to give $\SI{1}{\gram}$ of Compound (1) ($\SI{51}{\percent}$ yield) as a brown solid. $^1$H NMR ($\SI{400}{\mega\hertz}$, $\mathrm{CD_3COCD_3}$, $\delta$, ppm): 9.60 (bs, 2H), 8.19 (bs, 1H), 7.02 (d, $J = \SI{8.3}{\hertz}$, 2H), 6.73 (d, $J = \SI{8.3}{\hertz}$ 2H), 6.66-6.65 (m, 2H), 5.96 (q, 2H), 5.73-5.71 (m, 2H), 5.34 (s, 1H). $^{13}$C NMR ($\SI{100}{\mega\hertz}$, $\mathrm{CD_3COCD_3}$, $\delta$, ppm): 155.8, 134.5, 133.7, 129.3, 116.7, 114.8, 107.1, 106.3, 43.3.

\subsection{Synthesis of Compound 2}

Compound (2), 4,4-Difluoro-8-(4-hydroxyphenyl)-4-bora-3a,4a-diaza-s-indacene  was synthesized following a procedure in the literature \cite{refMexico}. Compound (1) ($\SI{1.0}{\gram}$, $\SI{4.20}{\milli\mole}$) and DDQ ($\SI{1.14}{\gram}$, $\SI{5.04}{\milli\mole}$) were dissolved in THF ($\SI{30}{\milli\liter}$). The reaction mixture was stirred at room temperature for two h. After of this time, triethylamine ($\SI{8.77}{\milli\liter}$, $\SI{62.95}{\milli\mole}$) was added and after 10 minutes $\mathrm{BF_3OEt_2}$ ($\SI{10.36}{\milli\liter}$, $\SI{83.93}{\milli\mole}$) was added dropwise. The reaction mixture was stirred for 2 h and then washed with water and extracted with ethyl acetate. The organic phase was filtered over $\mathrm{Na_2SO_4}$, the solvent was removed under reduced pressure and the crude product was further purified by column chromatography over silica gel with hexane/ethyl acetate (7:3) as eluent to give $\SI{0.39}{\gram}$ of 2 ($\SI{35}{\percent}$ yield) as a red solid. $^1$H NMR ($\SI{400}{\mega\hertz}$, $\mathrm{CD_3COCD_3}$, $\delta$, ppm): 9.28 (s, 1H), 7.97 (bs, 2H), 7.60 (d, $J = \SI{8.72}{\hertz}$, 2H), 7.11-7.08 (m, 4H), 6.66 (d, $J = \SI{4.1}{\hertz}$, 2H). $^{13}$C NMR ($\SI{100}{\mega\hertz}$, $\mathrm{CD_3COCD_3}$, $\delta$, ppm): 161.6, 149.0, 144.1, 135.5, 133.8, 132.2, 125.9, 119.2, 116.6.

\subsection{Synthesis of BODIPY-2-OH (Compound 3)}

To solution of Compound (2) ($\SI{0.5}{\gram}$, $\SI{1.76}{\milli\mole}$) in THF anhydrum ($\SI{20}{\milli\liter}$) under $\mathrm{N_2}$ atmosphere, was added $\mathrm{NaH}$ ($\SI{46.5}{\milli\gram}$, $\SI{1.94}{\milli\mole}$) and the reaction mixture was stirred for 30 minutes followed by the addition of 3-bromopropanol ($\SI{318.0}{\milli\gram}$, $\SI{2.29}{\milli\mole}$). The reaction mixture was stirred for 4 h at room temperature and then quenched with aq. $\mathrm{NH_4Cl}$. The reaction was extracted with ethyl acetate and water. The organic phase was dried ($\mathrm{Na_2SO_4}$) and the solvent was evaporated under reduced pressure. Purification by column chromatography (hexane/EtOAc, 7:3) gave 4,4-Difluoro-8-(4-(3-hydroxypropoxy)phenyl)-4-bora-3a,4a-diaza-s-indacene or BODIPY-2-OH, Compound (3) ($\SI{80}{\percent}$ yield) as a red solid. FTIR-ATR ($\nu \, \SI{}{\per\centi\meter}$): 3330,3145, 3120, 2960, 2938, 1729, 1541, 1384, 1249, 1184, 1119, 1071, 1054, 971, 841, 741, 707.  $^1$H NMR ($\SI{300}{\mega\hertz}$, $\mathrm{CDCl_3}$, $\delta$, ppm): 7.91 (bs, 2H), 7.52 (d, $J = \SI{8.7}{\hertz}$, 2H), 7.04 (d, $J = \SI{8.7}{\hertz}$, 2H), 6.96 (d, $J = \SI{3.8}{\hertz}$, 2H), 6.54 (d, $J = \SI{3.8}{\hertz}$, 2H), 4.22 (t, $J = \SI{5.9}{\hertz}$, 2H), 3.90 (t, $J = \SI{5.9}{\hertz}$, 2H), 2.10 (q, $J = \SI{5.9}{\hertz}$, 2H). $^{13}$C ($\SI{75}{\mega\hertz}$, $\mathrm{CDCl_3}$, $\delta$, ppm): 161.5, 147.5, 143.5, 134.9, 132.6, 131.5, 126.5, 118.4, 114.7, 66.7, 60.0, 32.0. $^{11}$B ($\SI{160.4}{\mega\hertz}$, $\mathrm{CDCl_3}$, $\delta$, ppm): -0.67 (t, $J_\mathrm{B-F}$= 28.9 Hz). $^{19}$F ($\SI{282.4}{\mega\hertz}$, CDCl$_{3}$, $\delta$, ppm): -144.8 (q, $J_\text{B-F} = \SI{28.9}{\hertz}$). HRMS (DART) m/z Calcd. for $\mathrm{C_{18}H_{17}BF_2N_2O_2} + \mathrm{H^+} = 343.14294$ found 343.14391 (2.82 ppm).

\begin{figure*}[!htb]
    \centering
    \includegraphics{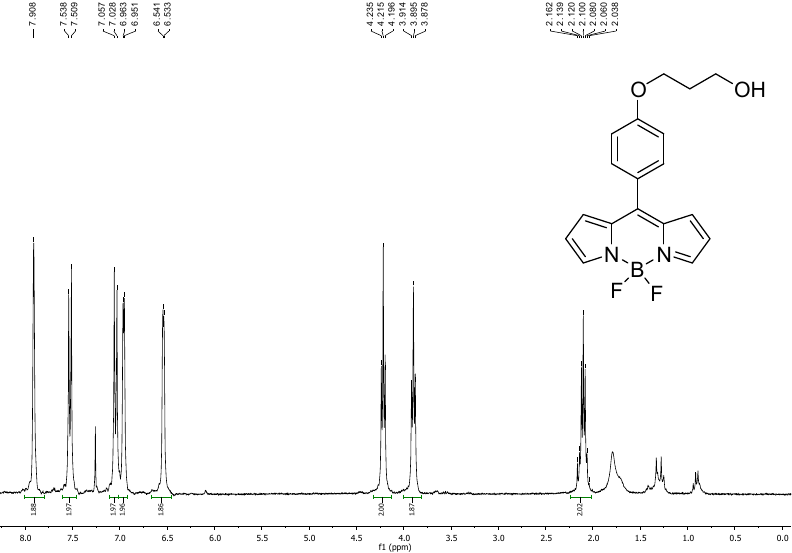}
    \caption{$\mathrm{^{1}H}$ NMR spectrum of compound 3 at $\SI{75}{\mega\hertz}$ in $\mathrm{CDCl_3}$.}
    \label{spectrum1}
\end{figure*}

\begin{figure*}[!htb]
    \centering
    \includegraphics{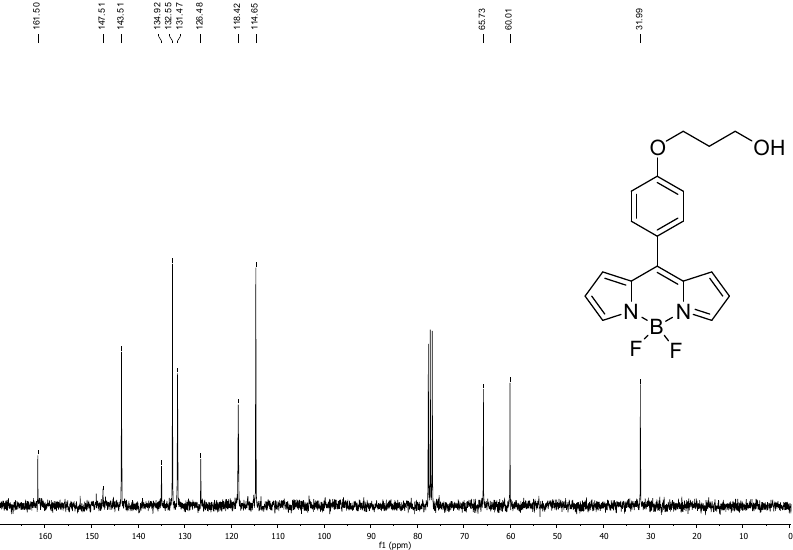}
    \caption{$\mathrm{^{13}C}$ NMR spectrum of compound 3 at $\SI{75}{\mega\hertz}$ in $\mathrm{CDCl_3}$.}
    \label{spectrum2}
\end{figure*}

\begin{figure*}[!htb]
    \centering
    \includegraphics{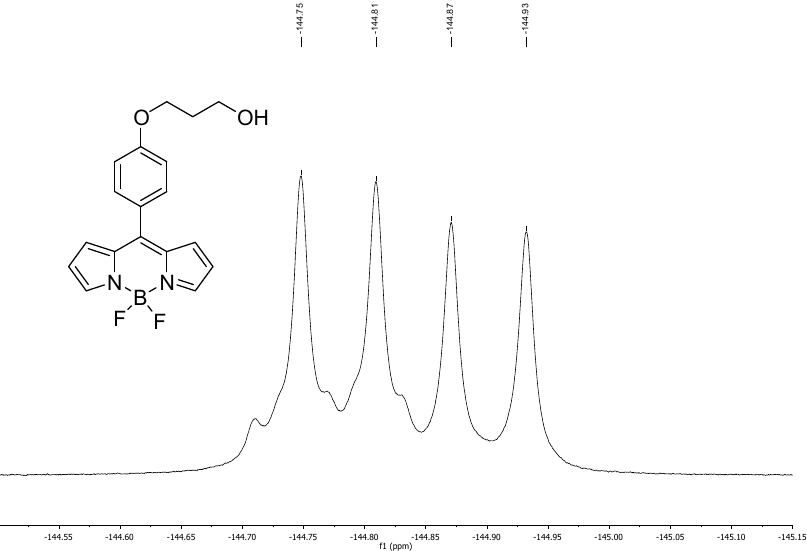}
    \caption{$\mathrm{^{19}F}$ NMR spectrum of compound 3 at $\SI{282.4}{\mega\hertz}$ in $\mathrm{CDCl_3}$.}
    \label{spectrum3}
\end{figure*}

\begin{figure*}[!htb]
    \centering
    \includegraphics{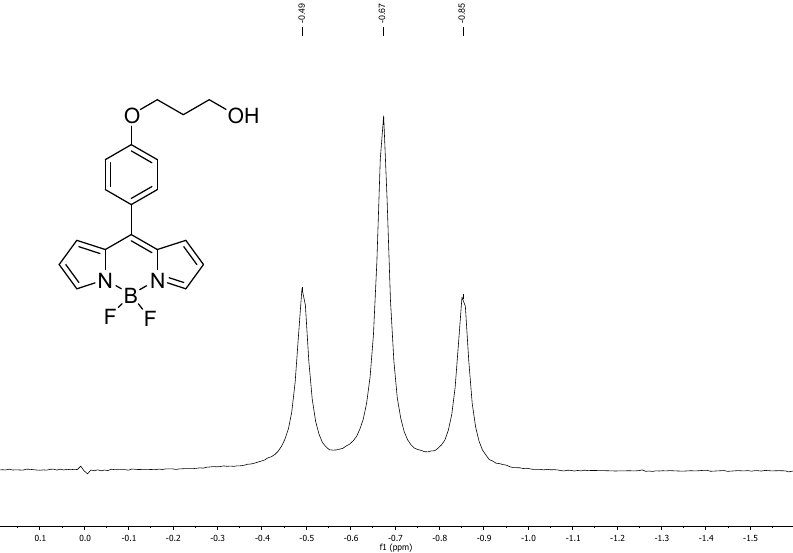}
    \caption{$\mathrm{^{11}B}$ NMR spectrum of compound 3 at $\SI{160.4}{\mega\hertz}$ in $\mathrm{CDCl_3}$.}
    \label{spectrum4}
\end{figure*}

$\mathrm{^{1}H}$, $\mathrm{^{13}C}$, $\mathrm{^{19}F}$ and $\mathrm{^{11}B}$ NMR spectra of compound (3) are respectively displayed on Figures~\ref{spectrum1},~\ref{spectrum2},~\ref{spectrum3} and~\ref{spectrum4}.

\newpage
\phantom{.}

\section{Viscosity-Composition calibration curves}

\subsection{Viscosity-Glycerol Volume Fraction Calibration Curve}

\begin{figure}[!htb]
    \includegraphics[scale=0.4]{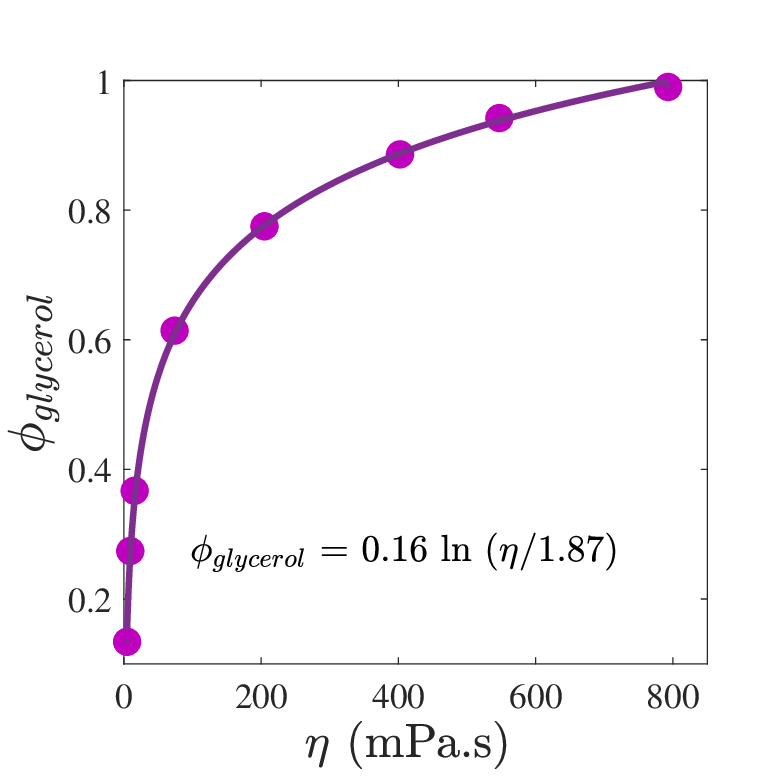}
    \caption{Glycerol volume fraction $\phi$ as a function of the viscosity $\eta$ of the DMSO-glycerol mixture. The continuous line corresponds to Eq.~\eqref{eq:phi_to_eta_calib} with $a = 0.16$ and $\eta_0 = \SI{1.87}{\milli\pascal\second}$.}
    \label{etaToVolFrac_CF}
\end{figure}

As described in the main text, the viscosity $\eta$ of mixtures of various glycerol volume fraction $\phi$ was measured. These parameters can be related by a fitting equation:
\begin{equation}
    \phi = a \ln \left(\frac{\eta}{\eta_0}\right)
    \label{eq:phi_to_eta_calib}
\end{equation}
\noindent with $a=0.16$ and $\eta_0=\SI{1.87}{\milli\pascal\second}$. Experimental measurements and fitting curve are displayed in Figure~\ref{etaToVolFrac_CF}.

\subsection{\large Glycerol Concentration-Viscosity Calibration Curve}

\begin{figure}[H]
     \centering
    \includegraphics[scale=0.4]{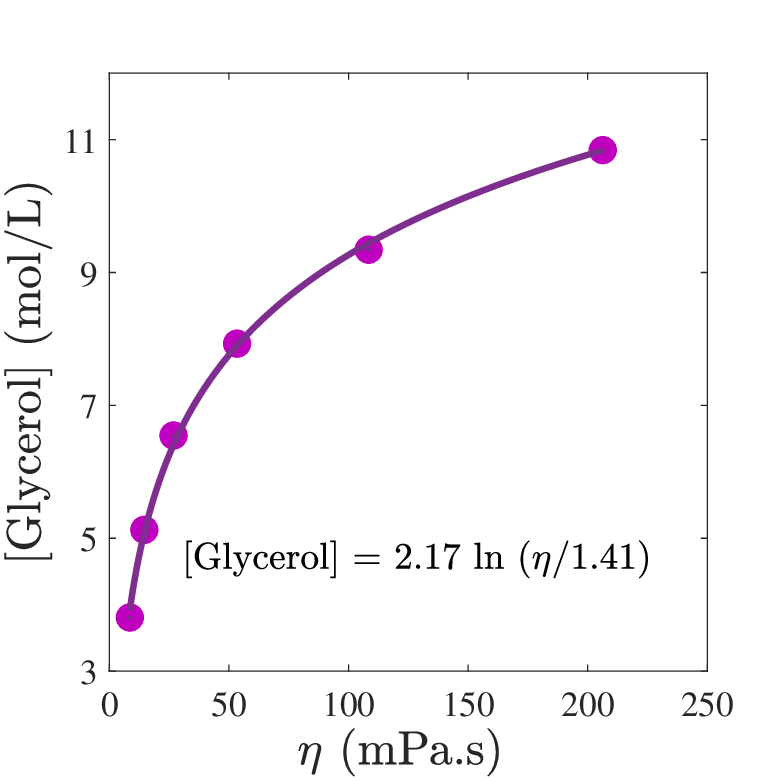}
\caption{Glycerol concentration $[\mathrm{Glycerol}]$ as a function of the viscosity $\eta$ of the DMSO-glycerol mixture. The continuous line corresponds to Eq.~\eqref{eq:conc_to_eta_calib} with $a'=\SI{2.17}{\mole\per\liter}$ and $\eta_0'=\SI{1.41}{\milli\pascal\second}$.}
\label{etaToConc_MM2}
\end{figure}

Instead of volume fraction, it can also be useful to measure glycerol concentration $[\mathrm{Glycerol}]$. It can be related to viscosity through:
\begin{equation}
    [\mathrm{Glycerol}] = a' \ln \left(\frac{\eta}{\eta'_0}\right)
    \label{eq:conc_to_eta_calib}
\end{equation}
\noindent with $a'=\SI{2.17}{\mole\per\liter}$ and $\eta_0'=\SI{1.41}{\milli\pascal\second}$. Experimental measurements and fitting curve are displayed in Figure~\ref{etaToConc_MM2}.

\providecommand{\latin}[1]{#1}
\makeatletter
\providecommand{\doi}
  {\begingroup\let\do\@makeother\dospecials
  \catcode`\{=1 \catcode`\}=2 \doi@aux}
\providecommand{\doi@aux}[1]{\endgroup\texttt{#1}}
\makeatother
\providecommand*\mcitethebibliography{\thebibliography}
\csname @ifundefined\endcsname{endmcitethebibliography}
  {\let\endmcitethebibliography\endthebibliography}{}

\end{document}